\documentclass[11pt]{article}
\usepackage{amssymb}
\usepackage{amsfonts}
\usepackage{graphicx}
\usepackage{amsmath}
\usepackage{hyperref}

\makeatletter \@addtoreset{equation}{section} \makeatother

\newtheorem{theorem}{Theorem}
\newtheorem{lemma}{Lemma}

\newtheorem{remark}{Remark}

\newtheorem{proposition}{Proposition}

\begin{document}

\title{Orthogonal and symplectic matrix models: universality and other properties.}
\author{ M. Shcherbina
\\
Institute for Low Temperature Physics Ukr.Ac.Sci, Kharkov, Ukraine.\\
E-mail:shcherbi@ilt.kharkov.ua
}

\date{}

\maketitle

\begin{abstract}
We study orthogonal and symplectic matrix models with polynomial potentials and multi interval
supports of the equilibrium measure. For these models we find the bounds (similar to the case of
hermitian matrix models) for the rate of convergence of linear eigenvalue
statistics and for the variance of linear eigenvalue
statistics and find the logarithms of partition functions up to the order
$O(1)$. We prove also  universality of  local eigenvalue statistics in the bulk.
\end{abstract}

\section{Introduction and main results}\label{s:1}

In this paper we consider ensembles of random matrices, whose joint eigenvalues distribution
is
\begin{equation}\label{p(la)}
p_{n,\beta}(\lambda_1,...,\lambda_{n})=Q_{n,\beta}^{-1}[V]\prod_{i=1}^n
e^{-n\beta V(\lambda_i)/2}\prod_{1\le i<j\le
n}|\lambda_i-\lambda_j|^\beta=Q_{n,\beta}^{-1}e^{\beta H(\lambda_1,\dots,\lambda_n)/2},
\end{equation}
where the function $H$, which we call Hamiltonian to stress the analogy with statistical mechanics,
and the normalizing constant $Q_{n,\beta}[V]$ have the form
\begin{eqnarray}
&&H(\lambda_1,\dots,\lambda_n)=-n\sum_{i=1}^n
V(\lambda_i)+\sum_{i\not=j}\log|\lambda_i-\lambda_j|,\notag\\
&&Q_{n,\beta}[V]=\int e^{\beta H(\lambda_1,\dots,\lambda_n)/2}d\lambda_1\dots d\lambda_{n}.
\label{Q}\end{eqnarray}
The function $V$, called the potential, is a real valued H\"{o}lder function satisfying the condition
\begin{equation}\label{condV}
V(\lambda )\ge 2(1+\epsilon )\log(1+ |\lambda |).
\end{equation}
  This distribution can be considered
for any $\beta>0$, but the cases $\beta=1,2,4$ are especially important, since they
correspond  to real symmetric, hermitian, and symplectic
matrix models respectively.

 We will consider also the marginal densities of (\ref{p(la)}) (correlation functions)
\begin{equation}\label{p_nl}
p^{(n)}_{l,\beta}(\lambda_1,...,\lambda_l)=
\int_{\mathbb{R}^{n-l}} p_{n,\beta}(\lambda_1,...\lambda_l,\lambda_{l+1},...,\lambda_{n})
d\lambda_{l+1}...d\lambda_{n},
\end{equation}
and denote
\begin{equation}\label{E}
    \mathbf{E}_\beta\{(\dots)\}=\int(\dots)p_{n,\beta}(\lambda_1,...,\lambda_{n})d\lambda_1,\dots
    d\lambda_{n}.
\end{equation}
It is known (see \cite{BPS:95,Jo:98}) that if $V'$ is a H\"{o}lder function, then
the first marginal density $p^{(n)}_{1,\beta}$ converges weakly to the function
$\rho$ (equilibrium density) with a compact support
$\sigma$. The density $\rho$  maximizes the
functional, defined on the class $\mathcal{M}_1$ of positive unit measures on $\mathbb{R}$
\begin{equation}\label{E_V}
\mathcal{E}_V(\rho)=\max_{m\in\mathcal{M}_1}\bigg\{
L[\,dm,dm]-\int V(\lambda)m(d\lambda)\bigg\}=\mathcal{E}[V],
\end{equation}
where we denote
\begin{equation}\label{L[,]}
L[\,dm,dm]=\int\log|\lambda-\mu|dm(\lambda) dm(\mu),\quad
L[f,g]=\int\log|\lambda-\mu|f(\lambda)g(\mu)d\lambda d\mu.
\end{equation}
The support $\sigma$ and the density $\rho$ are uniquely defined by the conditions:
\begin{equation}\label{cond_rho}\begin{array}{l}\displaystyle
v(\lambda ):=2\int \log |\mu -\lambda |\rho (\mu )d\mu -V(\lambda )=\sup v(\lambda):=v^*,\quad\lambda\in\sigma\\
v(\lambda )\le \sup v(\lambda),\quad \lambda\not\in\sigma,\hskip
2cm\sigma=\hbox{supp}\{\rho\}.
\end{array}\end{equation}

For $\beta=2$ it is well known (see \cite{Me:91}) that all correlation functions (\ref{p_nl}) can be represented as
\begin{equation} \label{p_k=}
p^{(n)}_{l,\beta}(\lambda_1,...,\lambda_l)=
\frac{(n-l)!}{n!}\det \{K_{n,2}(\lambda
_{j},\lambda _{k})\}_{j,k=1}^{l},
\end{equation}%
where
\begin{equation}\label{K_2}
K_{n,2}(\lambda ,\mu )=\sum_{l=0}^{n-1}\psi _{l}^{(n)}(\lambda )\psi
_{l}^{(n)}(\mu ).
\end{equation}%
This function is known as a reproducing kernel of the orthonormalized system
\begin{equation}\label{psi}
\psi _{l}^{(n)}(\lambda )=\exp \{-n V(\lambda )/2\}p_{l}^{(n)}(\lambda
),\;\,l=0,...,
\end{equation}
in which $\{p_{l}^{(n)}\}_{l=0}^n$ are orthogonal polynomials on $\mathbb{R}$
associated with the weight
$ w_{n}(\lambda )=e^{-n V(\lambda )}$, i.e.,
\begin{equation}\label{ortP}
\int p_{l}^{(n)}(\lambda )p_{m}^{(n)}(\lambda )w_{n}(\lambda )d\lambda
=\delta _{l,m}.
\end{equation}%
The orthogonal polynomial machinery, in particular, the Christoffel-Darboux formula and
Christoffel function simplify considerably the studies of  marginal densities (\ref{p_nl}).
This allows to study the local eigenvalue statistics in many different cases: bulk of the spectrum,
edges of the spectrum, special points, etc. (see \cite{PS:97},
\cite{PS:07},\cite{DKMVZ:99},\cite{BI:03},\cite{C-K:06},\cite{S:08},\cite{Mi-McL:08}).

For $\beta=1,4$ the situation is more complicated. It was shown in \cite{Tr-Wi:98} that all
correlation functions can be expressed in terms of some matrix kernels (see (\ref{Kn1}) --
(\ref{D,M}) below). But the representation is less convenient than (\ref{p_k=}) -- (\ref{K_2}).
It makes difficult the problems, which for $\beta=2$ are just simple exercises. For example, the bound for the
variance of linear eigenvalue statistics (\ref{lst}) for $\beta=1,4$ till now was known only for one interval
 $\sigma$ (see \cite{Jo:98}), while for $\beta=2$ it is a trivial corollary of the
Christoffel-Darboux formula for any $\sigma$.

The matrix kernels for $\beta=1,4$ have the form
\begin{align}\label{Kn1}
     K_{n,1} (\lambda,\mu)&:=
     \begin{pmatrix}
          S_{n,1}(\lambda,\mu) & -\frac{\partial}{\partial \mu} S_{n,1}(\lambda,\mu)\\
          (\epsilon S_{n,1}) (\lambda,\mu)-\epsilon (\lambda-\mu) & S_{n,1}(\mu,\lambda)
     \end{pmatrix}\ \textrm{for}\ \beta=1, n\ \textrm{even,}\\
     \label{Kn4}
     K_{n,4}(\lambda,\mu) &:=
     \begin{pmatrix}
          S_{n,4}(\lambda,\mu) & -\frac{\partial}{\partial \mu} S_{n,4}(\lambda,\mu)\\
          (\epsilon S_{n,4}) (\lambda,\mu) & S_{n,4}(\mu,\lambda)
     \end{pmatrix}\ \textrm{for}\ \beta=4,
\end{align}
where
\begin{align}\label{Sn1}
     S_{n,1}(\lambda,\mu) &= -\sum_{j,k=0}^{n-1} \psi_j^{(n)}(\lambda)
     (M_n^{(n)})_{jk}^{-1} (\epsilon \psi_k^{(n)})(\mu),\\
     \label{Sn4}
     S_{n/2,4}(\lambda,\mu) &= -\sum_{j,k=0}^{n-1} (\psi_j^{(n)})'(\lambda)
     (D_n^{(n)})_{jk}^{-1}  \psi_k^{(n)}(\mu),
\end{align}
$\epsilon (\lambda)=\frac{1}{2} \hbox{sgn} (\lambda)$,  $\hbox{sgn}$ denotes the standard signum
function,
\[(\epsilon f)(\lambda):=\int_{\mathbb R}\epsilon (\lambda-\mu)f(\mu)\, d\mu
d\lambda'.\]
$D^{(n)}_n$ and $M^{(n)}_n$  in (\ref{Sn1}) and (\ref{Sn4}) are the left top corner $n\times n$ blocks of the
semi-infinite matrices  that correspond to the differentiation
operator and to some integration operator respectively.
\begin{align}\label{D,M}
     D^{(n)}_\infty &:= \left(\left(\psi_j^{(n)}\right)',\psi_k^{(n)}\right)_{j,k\ge 0},
     \quad D^{(n)}_n=\{D^{(n)}_{jk}\}_{j,k=0}^n,\\
     M_\infty^{(n)} &:= \left (\epsilon \psi_j^{(n)},\psi_k^{(n)}\right)_{j,k\ge 0},\
     \quad M^{(n)}_n=\{M^{(n)}_{jk}\}_{j,k=0}^n.
\notag\end{align}
Both matrices $D_\infty^{(n)}$ and $M_\infty^{(n)}$ are skew-symmetric, and since
 $\epsilon (\psi_j^{(n)})'=\psi_j^{(n)}$, we have for any $j,l\ge 0$ that
\begin{equation*}
     \delta_{jl}=(\epsilon (\psi_j^{(n)})',\psi_l)=\sum_{k=0}^\infty (D_\infty^{(n)})_{jk}(M_\infty^{(n)})_{kl}
\quad \Longleftrightarrow\quad D_\infty^{(n)} M_\infty^{(n)}=1 =M_\infty^{(n)} D_\infty^{(n)}.
\end{equation*}
It was observed  in \cite{Wi:99}
that, if  $V$ is a rational function,  in particular, a polynomial of degree $2m$,
then the kernels $S_{n,1},S_{n,4}$ can be written as
\begin{align}\label{Sn.1}
     S_{n,1}(\lambda,\mu) &= K_{n,2}(\lambda,\mu)+n\sum_{j,k=-(2m-1)}^{2m-1} F^{(1)}_{jk}\psi_{n+j}^{(n)}(\lambda)
     \epsilon\psi_{n+k}^{(n)}(\mu),\\
     S_{n/2,4}(\lambda,\mu)&= K_{n,2} (\lambda,\mu) +n\sum_{j,k=-(2m-1)}^{2m-1} F^{(4)}_{jk}\psi_{n+j}^{(n)}(\lambda)
     \epsilon\psi_{n+k}^{(n)}(\mu),
\notag\end{align}
where  $F_{jk}^{(1)}$, $F_{jk}^{(4)}$ can be expressed in terms of the matrix $T_n^{-1}$,
where $T_n$ is the
$(2m-1)\times (2m-1)$ block in the bottom right corner of $D_n^{(n)}M_n^{(n)} $, i.e.,
\begin{equation}\label{T}
     (T_n)_{jk}:= (D_n^{(n)}M_n^{(n)})_{n-2m+j, n-2m+k},\quad\quad 1\le j, k\le 2m-1.
\end{equation}
The main technical obstacle to study the kernels $S_{n,1},S_{n,4}$ is the problem to prove
that $(T_n^{-1})_{jk}$  are bounded uniformly in $n$.
Till now this technical problem was solved only in a few
cases. In the papers \cite{De-G:07,De-G:07a}  the case $V(\lambda)=\lambda^{2m}(1+o(1))$
   (in  our notations) was studied and the problem of invertibility of $T_n$ was solved
   by computing the entries of $T_n$ explicitly. Similar method
  was used in \cite{DGKV:07} to prove bulk and edge universality (including the case
of hard edge) for the Laguerre type ensembles with monomial $V$.
 In the paper \cite{St1} the problem of invertibility of $T_n$ was solved also by computing the entries of $T_n$
for  $V$ being an even quatric polynomial.
In \cite{S:08,S:09} similar problem was solved without explicit computation of the entries of
$T_n$, it was shown that for any real
analytic $V$  with one interval support of the equilibrium density
$(M_n^{(n)})^{-1}$ is uniformly bounded in the operator norm. This
allowed us to prove bulk and edge universality  for $\beta=1$ in the one interval case.

But there is also a possibility to prove that $T_n$ is invertible  with
 another technique. As a by product of the calculation in \cite{Tr-Wi:98} one also obtains
relations between the partition
functions $Q_{n,\beta}$ and the determinants of $M_n^{(n)}$ and $D_n^{(n)}$:
\begin{equation*}
     \det M_n^{(n)} =\left(\frac{Q_{n,1}\Gamma_n}{n!2^{n/2}}\right)^2\quad ,\quad
     \det D^{(n)}_n =\left(\frac{Q_{n/2,4}\Gamma_n}{(n/2)!2^{n/2}}\right)^2,
\end{equation*}
where $\Gamma_n:= \prod_{j=0}^{n-1} \gamma_j^{(n)}$, and $\gamma_j^{(n)}$ is the leading coefficient
of $p^{(n)}_j(\lambda)$ of (\ref{ortP}). It is also known (see \cite{Me:91}) that $Q_{n,2}=\Gamma_n^2/n!$. Since $ D_\infty^{(n)}M_\infty^{(n)}=1$ and $(D_\infty^{(n)})_{jk}=0$ for
$|j-k|>2m-1$ (see (\ref{D_j,k})), we have $D_n^{(n)}M_n^{(n)} =1 +\Delta_n$ with $\Delta_n$ being zero except
for the bottom $2m-1$ rows, and we arrive at a formula,
first observed  in \cite{St1}:
\begin{equation}\label{detT}
     \det (T_n)= \det (D_n^{(n)}M_n^{(n)}) =\left(\frac{Q_{n,1}Q_{n/2,4}}{Q_{n,2}(n/2)!2^n}\right)^2.
\end{equation}
Hence  to control $\det (T_n)$, it suffices to control $\log Q_{n,\beta}$ for $\beta=1,2,4$ up to the order $O(1)$.
In the paper \cite{KS:10} the corresponding expansion of $\log Q_{n,\beta}$ was constructed by
using some generalization of the method of \cite{Jo:98}. The original method was proposed to
study the fluctuations of  linear eigenvalue statistics
\begin{equation}\label{lst}
   \mathcal{N}_n[\varphi]=\sum_{i=1}^n \varphi(\lambda_i),
\end{equation}
in particular, to control the expectation and the variance of $n^{-1}\mathcal{N}_n[\varphi]$ up to
the terms $O(n^{-2})$ for any $\beta$, but only in the case of one interval support of the
equilibrium measure  and   polynomial $V$, satisfying some additional assumption.
 The method was used also to prove CLT for  fluctuations of $\mathcal{N}_n[\varphi]$.
In the paper \cite{KS:10} the method of \cite{Jo:98} was simplified, that allowed
to generalize it  on the case of real analytic $V$
with one interval support of $\rho$, without any other assumption. Unfortunately, there is no hope to
generalize the method of \cite{Jo:98, KS:10} on the case of multi interval support $\sigma$ directly,
because the method is based on
 solving of some integral equation  (see Eq.(\ref{eq_6}) below) which  is not uniquely
solvable in the case of multi interval support.

In the present paper the problem to control $ Q_{n,\beta}[V]$ for $\beta=1,2,4$ is solved in a
little bit different way. We prove that for analytical potential $V$ with $q$-interval support
$\sigma$ of the equilibrium density  $ Q_{n,\beta}$ can be factorized to a product of
$Q_{k_\alpha^*,\beta}[V_a^{(\alpha)}]$, $\alpha=1,\dots, q$, where $k_\alpha^*\sim \mu^*_\alpha n$
(see (\ref{mu^*})),  and the "effective potentials" $V^{(a)}_\alpha$  (see (\ref{V^a})) are defined
in terms of $\sigma$, $V$ and $\rho$.

 To be more precise let us formulate our main conditions.

\medskip \noindent \textbf{Condition C1. }\textit{$V$ is a polynomial of degree $2m$ with positive leading
coefficient, and the support
of its equilibrium measure is}
\begin{equation}\label{sigma}
    \sigma=\bigcup_{\alpha=1}^q\sigma_\alpha,\quad \sigma_\alpha=[E_{2\alpha-1},E_{2\alpha}]
\end{equation}
\medskip \noindent \textbf{Condition C2. }\textit{ The equilibrium density $\rho$ can be represented in the form
\begin{equation}\label{rho}
    \rho(\lambda)=\frac{1}{2\pi}P(\lambda)\Im X^{1/2}(\lambda+i0),\quad
    \inf_{\lambda\in\sigma}|P(\lambda)|>0,
\end{equation}
where
\begin{equation}\label{X}
    X(z)= \prod_{\alpha=1}^{2q}(z-E_\alpha),
    \end{equation}
and we choose a branch of $X^{1/2}(z)$ such that $X^{1/2}(z)\sim z^q$, as $z\to+\infty$.
Moreover, the function $v$ defined by (\ref{cond_rho})
attains its maximum only if $\lambda $ belongs to  $\sigma $. }

\begin{remark} It is known (see, e.g., \cite{APS:01}) that for analytic $V$
the equilibrium density $\rho$
 has the form (\ref{rho}) -- (\ref{X}) with $P\ge 0$.
 The function $P$ in
(\ref{rho}) is  analytic and  can be represented in the form
\begin{equation}\label{P}
    P(z)=\frac{1}{2\pi i}\oint_\mathcal{L}\frac{V'(z)-V'(\zeta)}{(z-\zeta) X^{1/2}(\zeta)}d\zeta.
    \end{equation}
Hence  condition C2 means that $\rho$ has no zeros in the internal points of $\sigma$ and behaves like
square root near the edge points. This behavior of $V$ is usually called generic.
\end{remark}
We will use also the notations
\begin{eqnarray}\label{s_e}
&&\sigma_\varepsilon=\bigcup_{\alpha=1}^q\sigma_{\alpha,\varepsilon},\quad
\sigma_{\alpha,\varepsilon}=[E_{2\alpha-1}-\varepsilon,E_{2\alpha}+\varepsilon],\\
&&\mathrm{dist\,}\{\sigma_{\alpha,\varepsilon},\sigma_{\alpha',\varepsilon}\}
>\delta>0,\quad \alpha\not=\alpha'.
\notag\end{eqnarray}

The first result of the paper is the theorem which allows us to control $\log Q_{n,\beta}$
in the one interval case up to  $O(1)$ terms. Since the paper \cite{BPS:95} it is known that
\begin{equation*}
\log Q_{n,\beta}[V]=\beta n^2\mathcal{E}[V]/2+O(n\log n).
\end{equation*}
But, as it was discussed above, for many problems it is important to control the next terms of asymptotic expansion
of $\log Q_{n,\beta}$ (see  also the discussion in \cite{Er-McL:02}, where the expansion in $n^{-1}$
 was constructed for  $\beta=2$ and $V$ being
 a polynomial, close in a certain sense  to $V_0(\lambda)=\lambda^2/2$.)

We would like to note that  almost all assertions of Theorem \ref{t:2} below were obtained in \cite{KS:10}.
The difference  is
that here we need to control that the remainder bounds  are uniform in some parameter $\eta$, which we put  in front
of $V$ in $H$ (see (\ref{Q})).
Note also that it is
important for us that here $V$ may be non polynomial function analytic only in a
some open domain $\mathbf{D}\subset\mathbb{C}$ containing $\sigma$.

We will use below the Stieltjes transform, defined for an integrable function $p$ as
\begin{equation}\label{St_tr}
    g(z)=\int\frac{p(\lambda)d\lambda}{\lambda-z}.
\end{equation}
\begin{theorem}\label{t:2}
Let $V$ satisfy (\ref{condV}),  the  equilibrium
density $\rho$ (see (\ref{cond_rho}))   have  the form
(\ref{rho}) with $q=1$, and $\sigma=\mathrm{supp\,}\rho=[a,b]$.
Assume also that $V$ is  analytic  in the domain $\mathbf{D}\supset \sigma_\varepsilon$.
Consider the distribution (\ref{p(la)}) with $V$ replaced by $\eta V$.
Then there exists $\varepsilon_1>0$ such that
for any $\eta: |\eta-1|\le\varepsilon_1$ we have:

\noindent(i) The Stieltjes transform $g_n^{(\eta)}(z)$ (\ref{St_tr}) of the  first marginal
$p^{(n,\eta)}_{1,\beta}$ of (\ref{p(la)}) for $z$ such that $d(z):=\mathrm{dist\,}\{ z,\sigma_\varepsilon\}\ge n^{-1/6}\log n$ has the form
\begin{eqnarray}\label{t2.1}
&&g_n^{(\eta)}(z)=g_{\eta}(z)+n^{-1}u_{n,\eta}(z),
\\
&&u_{n,\eta}(z)=\left(\frac{2}{\beta}-1\right)\frac{1}{2\pi iX^{1/2}_{\eta}(z)}\oint_{\mathcal{L}}
\frac{g'_{\eta}(\zeta)d\zeta}{P_{\eta}(\zeta)(z-\zeta)}+n^{-1}O(d^{-10}( z)),
\notag\end{eqnarray}
where $g_{\eta}(z)$ is the Stieltjes transform of the equilibrium density $\rho_{\eta}$, which
maximizes $\mathcal{E}[\eta V]$ of (\ref{E_V}), $X_\eta$ of (\ref{X})
corresponds to the support $[a_{\eta},b_{\eta}]$
of $\rho_{\eta}$, and $P_{\eta}$ is defined by (\ref{P}) for  $\eta V$.
The contour $\mathcal{L}$ here is chosen sufficiently close to $[a_{\eta},b_{\eta}]$
  to have $z$ and all zeros of $P_{\eta}$ outside of it.
 The remainder bound   is uniform in $|\eta-1|\le\varepsilon_1$.
 Moreover,
  for any  $\varphi$ with bounded fifth derivative $\varphi^{(5)}$
\begin{eqnarray}\label{t2.1a}
\int \varphi(\lambda)\Big(p^{(n,\eta)}_{1,\beta}(\lambda)-\rho_\eta(\lambda)\Big)d\lambda&=&
n^{-1}O(||\varphi||_\infty+||\varphi^{(5)}||_\infty),\\
||\varphi||_\infty&:=&\sup_{\lambda\in\sigma_\varepsilon}|\varphi(\lambda)|.
\notag\end{eqnarray}

\noindent(ii) There exists an analytic in $\mathbf{D}\setminus\sigma_\varepsilon$ function $u^*$ such that
\begin{eqnarray}
&&u_{n,\eta}(z)-u_{n,1}(z)=(\eta-1)u^*(z)+O((\eta-1)^2)+O(n^{-1}),\quad|\Im z|>d.
\label{t2.2}\end{eqnarray}

\noindent(iii) If $Q_{n,\beta}^{(\eta)}$ is defined by (\ref{Q}) for $\eta V$, then
\begin{eqnarray}\notag
\log Q_{n,\beta}^{(\eta)}&=&\log Q_{n,\beta}^{*}+\frac{\beta n^2}{2}\mathcal{E}[\eta V]+\frac{3\beta n^2}{8}
+
n\Big(1-\frac{\beta}{2}\Big)\log\frac{d_\eta}{2}
\\&&-\frac{\beta n}{2 }\int_0^1dt\oint_{\mathcal{L}}
 (\eta V(z)-V_{\eta}^{(0)}(z))u_{n,\eta}(z,t)dz,\notag\\
 u_{n,\eta}(z,t)&=&\frac{(2/\beta-1)}{2\pi iX_{\eta}^{1/2}(z)}\oint_{\mathcal{L}_{d}}
\frac{g'_{\eta}(\zeta,t)d\zeta}{P_{\eta}(\zeta,t)(z-\zeta)}+O(n^{-1}),\hskip3cm
\label{log}\end{eqnarray}
where $\mathcal{E}[\eta V]$, $X_\eta$ and
$[a_\eta,b_\eta]$ are the same as in (ii),
\begin{eqnarray*}
&&V_{\eta}^{(0)}(z)=2(z-c_{\eta})^2/d_{\eta}^2,\quad
c_{\eta}=(a_{\eta}+b_{\eta})/2,\quad
d_{\eta}=(b_{\eta}-a_{\eta})/2,\\
&&P_{\eta}(\lambda,t)=tP_{\eta}(\lambda)+\frac{4(1-t)}{d_{\eta}^2},
\quad g_{\eta}(z,t)=tg_{\eta}(z)+\frac{2(1-t)}{d_{\eta}^2}(z-c_{\eta}-X^{1/2}_{\eta}(z)),
\end{eqnarray*}
$Q_{n,\beta}^{*}$ is defined by the Selberg formula
\begin{eqnarray}\label{Sielb}
Q_{n,\beta}^{*}=n!\left(\frac{n\beta}{2}\right)^{-\beta n^2/4-n(1-\beta/2)/2}(2\pi)^{n/2}\prod_{j=1}^{n}
\frac{\Gamma(\beta j/2)}{\Gamma(\beta /2)},\end{eqnarray}
and the remainder bound  is uniform in $|\eta-1|\le\varepsilon_1$.
\end{theorem}
The next theorem establishes some important properties of the symplectic and orthogonal matrix
models, in particular, it gives the bound for the rate of convergence of linear eigenvalue statistics and
 the bound for their variances. In order to formulate the theorem, we define  for even $n$
 \begin{equation}\label{mu^*}
    \mu^*_\alpha=\int_{\sigma_\alpha}\rho(\lambda)d\lambda, \quad
    k^*_\alpha:=[n\mu^*_\alpha]+d_\alpha,
\end{equation}
where $[x]$ means an integer part of $x$, and we have chosen $d_\alpha=0,\pm1,\pm2$ in a way which makes
$k^*_\alpha$ even and
\begin{equation}\label{sum_k}
\sum k^*_\alpha=n.\end{equation}
For each $\sigma_{\alpha,\varepsilon}$ we introduce the "effective potential"
\begin{eqnarray}\label{V^a}
  V^{(a)}_\alpha(\lambda)&=&\mathbf{1}_{\sigma_{\alpha,\varepsilon}}(\lambda)
\Big(V(\lambda)-2\int_{\sigma\setminus\sigma_\alpha}\log|\lambda-\mu|\rho(\mu)d\mu\Big),
\end{eqnarray}
and denote  $\Sigma^*$ the "cross energy"
\begin{eqnarray}\label{S*}\Sigma^*:&=&
 \sum_{\alpha\not=\alpha'}\int_{\sigma_\alpha}
 d\lambda\int_{\sigma_{\alpha'}}d\mu\log|\lambda-\mu|
 \rho(\lambda)\rho(\mu).
 \end{eqnarray}

\begin{theorem}\label{t:1} If potential $V$ satisfies conditions C1-C2 and $n$ is even, then
 the matrices $F^{(1)}$
and $F^{(4)}$ in  (\ref{Sn.1}) are
 bounded in the operator norm uniformly in $n$. Moreover, for any smooth $\varphi$
and $\beta=1,2,4$ we have
\begin{eqnarray}\label{c2.1}
&&\bigg|\int
\varphi(\lambda)\Big(p^{(n)}_{1,\beta}(\lambda)-\rho(\lambda)\Big)d\lambda\bigg|
\le\frac{C}{n}||\varphi'||_\infty,\\
&&\mathbf{E}_\beta\left\{ \Big| \mathcal{N}_n[\varphi]-\mathbf{E}_\beta\{
\mathcal{N}_n[\varphi]\}\Big|^2\right\}
\le C||\varphi'||^2_\infty,
\notag\end{eqnarray}
where $||.||_\infty$ is defined in (\ref{t2.1a}). The logarithm of the normalization constant
$Q_{n,\beta}[V]$ can be obtained up to $O(1)$ term from the representation
\begin{eqnarray}\label{fact}
\log( Q_{n,\beta}[V]/n!)&=&\sum_{\alpha=1}^q\log(Q_{k^*_\alpha,\beta}[V^{(a)}_\alpha]/k^*_\alpha!)
-\frac{\beta n^2}{2}\Sigma^*+O(1),
\end{eqnarray}
where $V^{(a)}_\alpha$ and $\Sigma^*$ are defined in (\ref{V^a}) and (\ref{S*}).

\end{theorem}
As it was mentioned above, Theorem \ref{t:1} together with some
asymptotic results  for orthogonal polynomials of \cite{DKMVZ:99} may be used  to prove universality
of  local eigenvalue statistics of  matrix models (\ref {p(la)}).
 In order to state our theorem on  bulk universality,
 we need some more notations.
Define
\begin{align*}
     K_\infty (t)&:= \frac{\sin \pi t}{\pi t},\\
     K_\infty^{(1)} (\xi, \eta)&:=
     \begin{pmatrix}
          K_\infty (\xi-\eta) & K'_\infty (\xi-\eta)\\
          \int^{\xi-\eta}_0 K_\infty (t)\, dt -\epsilon (\xi-\eta) & K_\infty (\eta-\xi )
     \end{pmatrix},\\
     K_\infty^{(4)} (\xi, \eta)&:=
     \begin{pmatrix}
          K_\infty (\xi-\eta) & K'_\infty (\xi-\eta)\\
          \int^{\xi-\eta}_0 K_\infty (t)\, dt & K_\infty (\eta-\xi )
     \end{pmatrix}.
\end{align*}
Furthermore, we denote for a $2\times 2$ matrix $A$ and $\lambda >0$
\begin{equation*}
      A^{(\lambda)} := \begin{pmatrix}
           \sqrt{\lambda}^{-1} & 0\\
           0&\sqrt{\lambda}
      \end{pmatrix}
      A \begin{pmatrix}
           \sqrt{\lambda} & 0\\
           0&\sqrt{\lambda}^{-1}
      \end{pmatrix}.
\end{equation*}

\begin{theorem} \label{Tt:2}
Let $V$  satisfy conditions C1--C2. Then we have for (even) $n\to \infty$,
$\lambda_0\in\mathbb R$ with $\rho (\lambda_0)>0$, and for $\beta \in \{1,4\}$ that
\begin{align*}
     &q_n^{-1} K_{n,1}^{(q_n)} \left(\lambda_0+{\xi}/{q_n},\lambda_0+{\eta}/{q_n} \right)
     =K_\infty^{(\beta)} (\xi,\eta)+ O(n^{-1/2}),\\
     &q_n^{-1} K_{n/2,4}^{(q_n)} \left(\lambda_0+{\xi}/{q_n},\lambda_0+{\eta}/{q_n} \right)
     =K_\infty^{(4)} (\xi,\eta)+ O(n^{-1/2}),
\end{align*}
where $q_n=n\rho(\lambda_0)$. The error bound is uniform for bounded $\xi$, $\eta$ and for
$\lambda_0$ contained in some compact subset of $\cup_{\alpha=1}^q(E_{2\alpha-1},E_{2\alpha})$.
\end{theorem}

It is an immediate consequence of Theorem \ref{Tt:2} that the corresponding rescaled $l$-point
correlation functions
\begin{equation*}
     p_{l,1}^{(n)} \left(\lambda_0+{\xi_1}/{q_n},\ldots ,\lambda_0+{\xi_l}/{q_n}\right),\quad
     p_{l,4}^{(n/2)} \left(\lambda_0+{\xi_1}/{q_n},\ldots ,\lambda_0+{\xi_l}/{q_n}\right)
\end{equation*}
converge for $n$ (even) $\to \infty$ to some limit that depends
on $\beta$ but not on the choice of $V$.

The paper is organized as follows. In Section \ref{s:2} we prove Theorem \ref{t:2}.
  In Section \ref{s:3} we prove Theorems
\ref{t:1} and \ref{Tt:2} modulo some  bounds which are
obtained in Section \ref{s:ap1}.

\section{Proof of Theorem \ref{t:2} }\label{s:2}

\textit{Proof of Theorem \ref{t:2}}.

\noindent(i). Take $n$-independent $\varepsilon$, small enough to
provide that $\sigma_\varepsilon\subset\mathbf{D}$. It is known (see e.g. \cite{BPS:95})
that if we replace in (\ref{p(la)}), (\ref{p_nl}), and  (\ref{E}) the integration over $\mathbb{R}$
by the integration $\sigma_\varepsilon$, then the new  marginal densities will differ from
the initial ones by the terms  $O(e^{-nc})$ with some $c$, depending on $\varepsilon$, but
independent of $n$. Since for our purposes it is more convenient to consider the
integration with respect to $\sigma_\varepsilon$, we assume from this moment that this
replacement is made, so everywhere below the integration without limits means the
integration over $\sigma_\varepsilon$.

Following  the idea of
\cite{Jo:98}, we will study a little bit modified form of the joint eigenvalue
distribution, than in (\ref{p(la)}). Namely, consider some real smooth function  $h(\lambda)$ and denote
\begin{equation}\label{V_eta}
V_{h,\eta}(\zeta)=\eta V(\zeta)+\frac{1}{n}h(\zeta).\end{equation}
Let $p_{n,\beta,h}$, $\mathbf{E}_{\beta,h}\{\dots\}$, $p^{(n,\eta)}_{l,\beta,h}$  be the distribution
density, the expectation, and the marginal densities defined by (\ref{p(la)}), (\ref{E}), and (\ref{p_nl})
with $V$ replaced by $V_{h,\eta}$.

By (\ref{p(la)})  the first  marginal density can be represented in the
form
\begin{equation}\label{r_rho}
p^{(n,\eta)}_{1,\beta,h}(\lambda)=Q_{n,\beta,h}^{-1}\int
e^{-n\beta V_{h,\eta}(\lambda)/2}\prod_{i=2}^n|\lambda-\lambda_i|^\beta
e^{-n\beta V_{h,\eta}(\lambda_i)/2}\prod_{2\le i<j\le
n}|\lambda_i-\lambda_j|^\beta d\lambda_2\dots d\lambda_n.
\end{equation}
Using the representation and integrating by parts,
we obtain
\begin{equation}\label{eq_1}
\int\frac{V'_{h,\eta}(\lambda)p^{(n,\eta)}_{1,\beta,h}(\lambda)}{z-\lambda} d\lambda=
\frac{2}{\beta n}\int\frac{p^{(n,\eta)}_{1,\beta,h}(\lambda)}{(z-\lambda)^2} d\lambda+
\frac{2(n-1)}{n}\int\frac{p^{(n,\eta)}_{2,\beta,h}(\lambda,\mu)d\lambda
d\mu}{(z-\lambda)(\lambda-\mu)}+O(e^{-nc}).
\end{equation}
Here  $O(e^{-nc})$ is the contribution of the integrated term. In fact all equations
below should contain $O(e^{-nc})$, but in order to simplify formula below we  omit it.

Since the function $p^{(n,\eta)}_{2,\beta,h}(\lambda,\mu)$ is symmetric with respect to
$\lambda,\mu$, we have
\[2\int\frac{p^{(n,\eta)}_{2,\beta,h}(\lambda,\mu)d\lambda d\mu}{(z-\lambda)(\lambda-\mu)}=
\int\frac{p^{(n,\eta)}_{2,\beta,h}(\lambda,\mu)d\lambda d\mu}{(z-\lambda)(\lambda-\mu)}+
\int\frac{p^{(n,\eta)}_{2,\beta,h}(\lambda,\mu)d\lambda d\mu}{(z-\mu)(\mu-\lambda)}=
\int\frac{p^{(n,\eta)}_{2,\beta,h}(\lambda,\mu)d\lambda d\mu}{(z-\lambda)(z-\mu)}.
\]
Hence  equation (\ref{eq_1}) can be written in the form
\begin{equation}\label{eq_2}
\int\frac{V'_{h,\eta}(\lambda)p^{(n,\eta)}_{1,\beta,h}(\lambda)}{z-\lambda} d\lambda=
\frac{2}{\beta n}\int\frac{p^{(n,\eta)}_{1,\beta,h}(\lambda)}{(z-\lambda)^2} d\lambda+
\frac{(n-1)}{n}\int\frac{p^{(n,\eta)}_{2,\beta,h}(\lambda,\mu)d\lambda
d\mu}{(z-\lambda)(z-\mu)}.
\end{equation}
Let us introduce notations:
\begin{multline}\label{delta}
    \delta^{(n,\eta)}_{\beta,h}(z)=n(n-1)\int\frac{p^{(n,\eta)}_{2,\beta,h}(\lambda,\mu)d\lambda
    d\mu}{(z-\lambda)(z-\mu)}-n^2\bigg(\int\frac{p^{(n,\eta)}_{1,\beta,h}(\lambda)d\lambda
    }{z-\lambda}\bigg)^2+n\int\frac{p^{(n,\eta)}_{1,\beta,h}(\lambda)}{(z-\lambda)^2}
    d\lambda\\=\int \frac{k^{(n,\eta)}_{\beta,h}(\lambda,\mu)d\lambda d\mu}{(z-\lambda)(z-\mu)},\hskip3cm
\end{multline}
where
\begin{equation}\label{k_n}
k^{(n,\eta)}_{\beta,h}(\lambda,\mu)=n(n-1)p^{(n,\eta)}_{2,\beta,h}(\lambda,\mu)-n^2p^{(n,\eta)}_{1,\beta,h}(\lambda)
p^{(n,\eta)}_{1,\beta,h}(\mu)+n\delta(\lambda-\mu)p^{(n,\eta)}_{1,\beta,h}(\lambda).
\end{equation}
Moreover, we denote
\begin{equation}\label{g_n}
  g^{(\eta)}_{n,\beta,h}(z)=  \int\frac{p^{(n,\eta)}_{1,\beta,h}(\lambda)d\lambda
    }{\lambda-z},\quad V(z,\lambda)=\frac{V'(z)-V'(\lambda)}{z-\lambda}.
\end{equation}
Then equation (\ref{eq_1}) takes the form
\begin{multline}\label{eq_3}
(g^{(\eta)}_{n,\beta,h}(z))^2+\eta V'(z)g^{(\eta)}_{n,\beta,h}(z)+
\eta\int V(z,\lambda)p^{(n,\eta)}_{1,\beta,h}(\lambda) d\lambda\\
=\frac{1}{n}\int\frac{h'(\lambda)p^{(n,\eta)}_{1,\beta,h}(\lambda)}{z-\lambda} d\lambda-
\frac{1}{n}\bigg(\frac{2}{\beta }-1\bigg)\int\frac{p^{(n,\eta)}_{1,\beta,h}(\lambda)}{(z-\lambda)^2} d\lambda
-\frac{1}{n^2}\delta_{n,\beta,h}(z).
\end{multline}
Using that $V(z,\zeta)$ is an analytic functions of $\zeta$ in $\mathbf{D}$, by the
Cauchy  theorem we obtain that
\begin{eqnarray*}
\int V(z,\lambda)p^{(n,\eta)}_{1,\beta,h}(\lambda) d\lambda=-\frac{1}{2\pi i}\oint_{\mathcal{L}}
V(z,\zeta)g^{(\eta)}_{n,\beta,h}d\zeta.\end{eqnarray*}
if $\mathcal{L}$ is encircling $\sigma_\varepsilon$, and  $z$  is outside of
$\mathcal{L}$. Thus (\ref{eq_3}) takes the form
\begin{multline}\label{eq_4}
(g^{(\eta)}_{n,\beta,h}(z))^2+\eta V'(z)g_{n,\beta,h}(z)-\frac{\eta}{2\pi i}\oint_{\mathcal{L}}
V(z,\zeta)g^{(\eta)}_{n,\beta,h}(\zeta)d\zeta\\=
\frac{1}{n}\int\frac{h'(\lambda)p^{(n,\eta)}_{1,\beta,h}(\lambda)}{z-\lambda}d\lambda-
\frac{1}{n}\bigg(\frac{2}{\beta }-1\bigg)\int\frac{p^{(n,\eta)}_{1,\beta,h}(\lambda)}
{(z-\lambda)^2}
d\lambda-\frac{1}{n^2}\delta_{n,\beta,h}(z).
\end{multline}
On the other hand, it is easy to show that $g_{\eta}$ satisfies the equation
\begin{equation}\label{eq_5}
g^2_{\eta}(z)+V'(z)g_{\eta}(z)+Q_{\eta}(z)=0,\quad Q(z)=-\frac{\eta}{2\pi i}\oint_{\mathcal{L}}
V(z,\zeta)g_{\eta}(\zeta)d\zeta.
\end{equation}
Hence
\[
g_{\eta}(z)=-\frac{\eta}{2}V'(z)+\frac{1}{2}\sqrt{\eta^2V'(z)^2-4Q_{\eta}(z)}.
\]
Using the inverse Stieltjes transform and comparing with (\ref{rho}), we get that
\begin{equation}\label{2g-V}
2g_{\eta}(z)+\eta V'(z)=P_{\eta}(z) X_{\eta}^{1/2}(z).\end{equation}
where $X_{\eta}(z)$ is defined by (\ref{X}) for q=1.

Write $g^{(\eta)}_{n,\beta,h}=g_{\eta}+n^{-1}u_{n,\eta}$.
Then, subtracting (\ref{eq_5}) from (\ref{eq_4}) and multiplying the result by $n$, we get
\begin{equation}\label{eq_6}
(2g_{\eta}(z)+\eta V'(z))u_{n,\eta}(z)-\frac{\eta}{2\pi i}\oint_{\mathcal{L}}
V(z,\zeta)u_{n,\eta}(\zeta)d\zeta=F(z),
\end{equation}
where
\begin{eqnarray}\label{F}
F(z)&=&
\int\frac{h'(\lambda)p^{(n,\eta)}_{1,\beta,h}(\lambda)}{z-\lambda}d\lambda
-\bigg(\frac{2}{\beta }-1\bigg)\left(g'_{\eta}(z)+\frac{1}{n}u_{n,\eta}'(z)\right)\\
&&-\frac{1}{n}u_{n,\eta}^2(z)-\frac{1}{n}\delta^{(n,\eta)}_{\beta,h}(z).
\notag\end{eqnarray}
Using (\ref{2g-V}), we obtain from (\ref{eq_6})
\begin{equation}\label{eq_7}
P_{\eta}(z)X_{\eta}^{1/2}(z)u_{n,\eta}(z)+\mathcal{Q}_n(z)=F(z),\quad
\mathcal{Q}_n(z)=-\frac{\eta}{2\pi i}\oint
V(z,\zeta)u_{n,\eta}(\zeta)d\zeta.
\end{equation}
Then, choosing  $\mathcal{L}$ sufficiently close to $[a_{\eta},b_{\eta}]$
to have all  zeros of $P(\zeta)$ and $z$ outside of $\mathcal{L}$, we
get
\begin{equation}\label{eq_8}
\frac{1}{2\pi i}\oint_{\mathcal{L}}\left(P_{\eta}(\zeta)X_{\eta}^{1/2}(\zeta)u_{n,\beta,h}(\zeta)
+\mathcal{Q}_n(\zeta)-F(\zeta)\right)\frac{d\zeta}{P_{\eta}(\zeta)(z-\zeta)}=0.
\end{equation}
Since  by definition (\ref{eq_7}) $\mathcal{Q}_n(\zeta)$ is  analytic  in
$\mathbf{D}$, and $z$ and all zeros of $P$ are outside of $\mathcal{L}$, the Cauchy
theorem yields
\[
\frac{1}{2\pi i}\oint_{\mathcal{L}}\frac{\mathcal{Q}_n(\zeta)d\zeta}{P_{\eta}(\zeta)(z-\zeta)}=0.
\]
Moreover,   since
\[u_{n,\eta}(z)=\frac{n}{z}\left(\int p^{(n,\eta)}_{1,\beta,h}(\lambda) d\lambda-
\int \rho_{\eta}(\lambda) d\lambda\right)+nO(z^{-2})=nO(z^{-2})
,\quad z\to \infty,
\]
we have
\begin{equation}\label{as_u}
X_{\eta}^{1/2}(z)u_{n,\eta}(z)=nO(z^{-1}).
\end{equation}
Then the Cauchy theorem yields
\[\frac{1}{2\pi
i}\oint_{\mathcal{L}}\frac{X_{\eta}^{1/2}(\zeta)u_{n,\eta}(\zeta)d\zeta}{(z-\zeta)}=X_{\eta}^{1/2}(z)u_{n,\eta}(z).
\]
Finally, we obtain from (\ref{eq_8})
\begin{equation}\label{eq_9}
u_{n,\eta}(z)=\frac{1}{2\pi iX_{\eta}^{1/2}(z)}\oint_{\mathcal{L}} \frac{F(\zeta)d\zeta}{P_{\eta}(\zeta)(z-\zeta)}.
\end{equation}
Set
\begin{equation}\label{L_d}
 d(z)=\textrm{dist}\{z,\sigma_\varepsilon\},
\end{equation}
Then for any $z:d(z)=d$ equation (\ref{eq_9}) implies
\begin{equation}\label{eq_10}
u_{n,\eta}(z)=\frac{F(z)}{X_{\eta}^{1/2}(z)P_{\eta}(z)}+
\frac{1}{2\pi iX_{\eta}^{1/2}(z)}\oint_{\mathcal{L}'
}\frac{F(\zeta)d\zeta}{P_{\eta}(\zeta)(\zeta-z)},
\end{equation}
where the contour $\mathcal{L}'$ contains $z$ but does not contain zeros of $P(z)$. It will be convenient
for us to take $\mathcal{L}'$ as far from $\sigma_\varepsilon$, as it is possible.
According to  \cite{BPS:95}, for any $\beta$ and $\eta$ we have the bounds:
\begin{equation}\label{b1_de}
    |\delta^{(n,\eta)}_{\beta,h}(z)|\le \frac{C n\log n}{d^4(z)},\quad
    \quad |u_{n,\eta}(z)|\le\frac{C n^{1/2}\log^{1/2} n}{d^2(z)},
     \quad |u_{n,\eta}'(z)|\le \frac{C n^{1/2}\log^{1/2} n}{d^3(z)},
\end{equation}
where $C$ is an absolute constant.
 Set
\[ M_n(d)=\sup_{z:d(z)\ge d}|u_{n,\eta}(z)|.\]
By (\ref{as_u}) and the maximum principle, there exists a point
$z:d(z)=d$ such that
\[
M_n(d)=|u_{n,\eta}(z)|.
\]
Then, using (\ref{eq_10}), the definition of $F$ (\ref{F}), and (\ref{b1_de}),
we obtain the inequality
\[M_n(d)\le \frac{M_n^2(d)}{C_1nd}+\frac{C_2\log n}{d^5},\quad C_2\le C_0(1+||h'||_{\infty})\]
with some  $C_0,C_1$ depending only on $P$ and $C$ of (\ref{b1_de}).
Solving the above  quadratic inequality, we get
\[
\left[\begin{array}{l}
M_n(d)\ge \frac{1}{2}\Big(C_1nd+\sqrt{C_1^2n^2d^2-4C_1C_2n\log n/d^4}\Big);\\
M_n(d)\le\frac{1}{2}\Big(C_1nd-\sqrt{C_1^2n^2d^2-4C_1C_2n\log n/d^4}\Big).
\end{array}\right.
\]
Since the first inequality contradicts to  (\ref{b1_de}), we conclude that for $d>n^{-1/6}\log n$ the second
inequality holds. Hence  we get
\begin{equation}\label{b_u}
|u_{n,\eta}(z)|\le C_0\log nd^{-5}(z)(1+||h'||_{\infty}),\quad d(z)>n^{-1/6}\log n.
\end{equation}
Note that the bound gives us that if we consider
$\varphi(\lambda)=\Re(\lambda-z)^{-1}$ or $\varphi(\lambda)=\Im(\lambda-z)^{-1}$ with $d(z)>n^{-1/6}\log n$, then
\begin{eqnarray}\label{b1_u}
n\bigg|\int\varphi(\lambda)(p^{(n,\eta)}_{1,\beta,h}(\lambda)-\rho_{\eta}(\lambda))d\lambda\bigg|
\le w_n ||\varphi^{(s)}||_{\infty}(1+||h'||_{\infty}),
\end{eqnarray}
where $\varphi^{(s)}$ is the $s$-th derivative of $\varphi$ (now we have $s=4$ in view of (\ref{b_u})) and
\[w_n =sC_0\log n\]
 We are going to use the following lemma, which is an analog of Lemma 3.11 of
 \cite{Jo:98}.
 \begin{lemma}\label{l:1}
 If (\ref{b1_u}) holds  for any real $h:||h'||_{\infty}<A$ ($A>1$) and some $\varphi$ such that
 $||\varphi^{(s)}||_{\infty}\ge||\varphi'||_{\infty}$, then there exists
 an absolute constant $C_*$ such
 that for any $||h'||_{\infty}<A/2$
\begin{equation}\label{bl_de}
\int k^{(n,\eta)}_{\beta,h}(\lambda,\mu)\varphi(\lambda)\varphi(\mu)d\lambda d\mu\le
C_*w_n^2(1+A)^2||\varphi^{(s)}||_{\infty}^2
\end{equation}
 \end{lemma}
 The lemma was proven in \cite{Jo:98}, but for reader's convenience  we
 give its proof here.

\textit{Proof of Lemma \ref{l:1}}. Without loss of generality assume that $w_n>1$.
 Using the method of \cite{Jo:98},
consider the function
\[ Z_n(t)=\mathbf{E}_{\eta,\beta,h}\left\{\exp\left[\frac{t}{2\widetilde w_n}\sum_{i=1}^n\Big(\varphi(\lambda_i)
-\int\varphi(\lambda)\rho_{\eta}(\lambda)d\lambda\Big)\right]\right\},\quad
 \widetilde w_n=w_n(1+A)||\varphi^{(s)}||_{\infty},
\]
where $\mathbf{E}_{\eta,\beta,h}\left\{.\right\}$ is defined in (\ref{E}) with $V$ replaced by $V_{\eta,h}$
of (\ref{V_eta}). It is easy to see that
\begin{equation}\label{logF}
\frac{d^2}{dt^2}\log Z_n(t)=(2w_n)^{-2}\mathbf{E}_{\beta,h+t\varphi/2\widetilde w_n}
\left\{\left(\sum_{i=1}^n(\varphi(\lambda_i)-\mathbf{E}_{\beta,h+t\varphi/2\widetilde w_n}\{\varphi(\lambda_i)\})
\right)^2\right\}\ge 0.
\end{equation}
Hence  in view of (\ref{b1_u}),
\begin{multline*}
\log Z_n(t)=\log Z_n(t)-\log Z_n(0)=\int_0^t\frac{d}{d\tau}\log Z_n(\tau)d\tau\le |t|\frac{d}{dt}\log
Z_n(t)\\=|t|(2\widetilde w_n)^{-1}\mathbf{E}_{\eta,\beta,h+t\varphi/2\widetilde w_n}
\left\{\sum_{i=1}^n\left(\varphi(\lambda_i)-\int\varphi(\lambda)\rho_{\eta}(\lambda)d\lambda\right)\right\}\\
=\frac{|t|n}{2\widetilde w_n}\int\varphi(\lambda)\left(p^{(n,\eta)}_{1,\beta,h+t\varphi/2\widetilde w_n}(\lambda)
-\rho_{\eta}(\lambda)\right)d\lambda\le |t|,\quad\quad t\in[-1,1].
\end{multline*}
Thus
\[Z_n(t)\le e^{|t|}\le 3,\quad t\in[-1,1],\]
and for any $t\in \mathbb{C}$, $|t|\le 1$
\begin{equation}\label{b_F}
   |Z_n(t)| \le Z_n(\Re t)<3.
\end{equation}
Then, by the Cauchy theorem, we have
\[ |Z_n'(t)|=\bigg|\frac{1}{2\pi}\oint_{|t'|=1}\frac{Z_n(t')dt'}{(t'-t)^2}\bigg|\le 12,
\quad |t|\le \frac{1}{2},\]
and therefore for $|t|\le\frac{1}{24}$
\[ |Z_n(t)|= |Z_n(0)-\int_{0}^tZ_n'(t)dt|\ge \frac{1}{2}.\]
Hence  $\log Z_n(t)$ is  analytic  for $|t|\le\frac{1}{24}$, and
using the above bounds we have
\[
\frac{d^2}{dt^2}\log Z_n(0)=
\frac{1}{2\pi i}\oint_{|t|=1/24}\frac{\log Z_n(t)}{t^3}dt\le C.
\]
Finally, in view of (\ref{logF}) we get
\[\int k_{n,\beta,h}(\lambda,\mu)\varphi(\lambda)\varphi(\mu)d\lambda d\mu=
\mathbf{E}_{\eta,\beta,h}
\left\{\left(\sum_{i=1}^n(\varphi(\lambda_i)-\mathbf{E}_{\eta,\beta,h}\{\varphi(\lambda_i)\})\right)^2\right\}
\le 4Cw_n^2.\]
$\square$

\medskip

 Let us come back to the proof of Theorem \ref{t:2}.
 Applying the lemma to $\varphi_z^{(1)}(\lambda)=\Re(z-\lambda)^{-1}$ and
 $\varphi_z^{(2)}(\lambda)=\Im(z-\lambda)^{-1}$ with $d(z)> n^{-1/6}\log n$ and using
 (\ref{b1_u}) for $||h'||_\infty\le 2$,  we obtain  for such $z$ (cf (\ref{b1_de}))
\begin{equation}\label{b2_de}
    |\delta^{(n,\eta)}_{\beta,h}(z)|\le C\log^2 nd^{-10}(z)
\end{equation}
Then, using this bound and (\ref{b_u}) in (\ref{eq_9}) and taking into account that
$n^{-1}\delta^{(n,\eta)}_{\beta,h}(z)\le d^{-4}(z)$ for
$d(z)> n^{-1/6}\log n$, we obtain that for $||h'||_\infty\le 1$
\begin{equation}\label{eq_11}
u_{n,\eta}(z)=\frac{(2/\beta-1)}{2\pi iX_{\eta}^{1/2}(z)}\oint_{\mathcal{L}_d}
\frac{g_{\eta}'(\zeta)d\zeta}{P_{\eta}(\zeta)(z-\zeta)}+O(d^{-5}(z)).
\end{equation}
Applying Lemma \ref{l:1} once more, we get  $\delta^{(n,\eta)}_{\beta,h}(z)\le Cd^{-10}(z)$.
Using the bound in (\ref{eq_10}) we prove (\ref{t2.1}).

\medskip

To prove (\ref{t2.1a}) consider the Poisson kernel
\begin{equation}\label{P_y}
\mathcal{P}_y(\lambda)=\frac{y}{\pi(y^2+\lambda^2)}.
\end{equation}
It is easy to see that for any integrable $\varphi$
\[(\mathcal{P}_y*\varphi)(\lambda)=\frac{1}{\pi}\Im\int\frac{\varphi(\mu)d\mu}{\mu-(\lambda+iy)}.\]
Hence we can use  (\ref{eq_11}) to prove that for $|y|\ge n^{-1/6}\log n$
\begin{equation}\label{P*u}
||\mathcal{P}_y*u_{n,\eta}||_2^2\le Cy^{-10},  \quad \hbox{for} \quad
u_{n,\eta}(\lambda):=n(p^{(n,\eta)}_{1,\beta}(\lambda)-\rho_\eta(\lambda)),
\end{equation}
where $||.||_2$ is the standard norm in $L_2(\mathbb{R})$.
Then we use the formula (see \cite{Jo:98}) valid for any $u\in L_2(\mathbb{R})$
\begin{equation}\label{Joh}
\int_0^\infty e^{-y}y^{2s-1}||\mathcal{P}_y*u||_2^2dy=\Gamma(2s)
\int_{\mathbb{R}}(1+2|\xi|)^{-2s}|\widehat{u}(\xi)|^2d\xi.
\end{equation}
The formula for $s=5$, the Parseval equation for the Fourier integral, and  the Schwarz inequality
yield
\begin{eqnarray*}
&&\hspace{-2cm}\int_{\mathbb{R}}\varphi(\lambda)u_{n,\eta}(\lambda)d\lambda=\frac{1}{2\pi}
\int_{\mathbb{R}}\widehat{\varphi}(\xi)\widehat{u}_{n,\eta}(\xi)d\xi\\
&\le&\frac{1}{2\pi}\left(\int_{\mathbb{R}}|\widehat{\varphi}(\xi)|^2(1+2|\xi|)^{2s}d\xi\right)^{1/2}
\left(\int_{\mathbb{R}}|\widehat{u}_{n,\eta}(\xi)|^2(1+2|\xi|)^{-2s}d\xi\right)^{1/2}\\
&\le& \frac{C((||\varphi||_{2}+||\varphi^{(5)}||_{2})}{\Gamma^{1/2}(2s)}\left(\int_0^\infty
e^{-y}y^{2s-1}||P_y*u_{n,\eta}||_2^2dy\right)^{1/2}.
\end{eqnarray*}
To estimate the last integral here we split it into two parts $|y|\ge n^{-1/6}\log n$ and $|y|< n^{-1/6}\log
n$. For the first integral we use  (\ref{P*u}) and for the second - (\ref{b1_de}).

(ii) To prove (\ref{t2.2}) we start from the relations
\begin{eqnarray}
&&a_{\eta}-a=(\eta-1)a_*+O((\eta-1)^2),\quad b_{\eta}-b=(\eta-1)b_*+O((\eta-1)^2),\notag\\
 &&P_{\eta}(z)-P(z)=(\eta-1)p(z)+O((\eta-1)^2),\notag\\
 &&g_{\eta}(z)-g(z)=(\eta-1)m(z)
 +O((\eta-1)^2|\Im z|^{-3}),\label{t2.2*}
\end{eqnarray}
where $a_*,b_*$ are some constant,  $p(z)$ is some analytic in $\mathbf{D}$ function, and $m(z)$ is
analytic outside $[a,b]$.
The first line here follows from the results of \cite{MK}.
The second line  follows from the first one, if we use the representation
(\ref{P}). The third line follows from the representation
\[g_{\eta}(z)=\frac{1}{2}(X^{1/2}_\eta(z)P_\eta(z)-\eta V'(z))\]
and the first two lines. Then (\ref{t2.2}) follows from (\ref{t2.1}), combined with (\ref{t2.2*}).

(iii) Consider  the functions $V_{\eta,t}$ of the form
\begin{equation}\label{V_t}
    V_{\eta,t}(\lambda)=t\eta V(\lambda)+(1-t)V^{(0)}_\eta(\lambda),
\end{equation}
where $V^{(0)}_\eta$ is defined in (\ref{log}).
Let $Q_{n,\beta}^{(\eta)}(t):=Q_{n,\beta}[V_{\eta,t}]$ be defined by (\ref{Q})
with $V$ replaced by $V_{\eta,t}$. Then, evidently, $Q_{n,\beta}^{(\eta)}(1)=Q_{n,\beta}[\eta V]$, and $Q_{n,\beta}(0)$
 corresponds to  $V^{(0)}_\eta$ (see (\ref{log})). Hence
\begin{align}\label{d_log_Q}
\frac{1}{n^2}\log Q_{n,\beta}^{(\eta)}(1)-\frac{1}{n^2}\log Q_{n,\beta}^{(\eta)}(0)&=
\frac{1}{n^2}\int_0^1dt\frac{d}{dt}\log Q_{n,\beta}^{(\eta)}(t)\\
&=-\frac{\beta}{2}\int_0^1dt\int d\lambda(\eta V(\lambda)-V^{(0)}_\eta(\lambda))p^{(n,\eta)}_{1,\beta}(\lambda;t),\notag
\end{align}
where $p^{(n,\eta)}_{1,\beta}(\lambda;t)$ is the first marginal density, corresponding to $V_{\eta,t}$.
Using (\ref{cond_rho}), one can check that  for the distribution
(\ref{p(la)}) with $V$ replaced by $V_t$ the equilibrium density $\rho_t$ has the form
\begin{equation}\label{rho_t}
\rho_t(\lambda)=t\rho_\eta(\lambda)+(1-t)\rho^{(0)}_\eta(\lambda),\quad
\rho^{(0)}_\eta(\lambda)=\frac{2X_\eta(\lambda)}{\pi d_\eta^2}
\end{equation}
with $X_\eta, d_\eta$ of (\ref{log}). Hence  using (\ref{t2.1})
for the last integral in (\ref{d_log_Q}), we get
\begin{eqnarray*}
\log Q_{n,\beta}[\eta V]&=&\log Q_{n,\beta}[V^{(0)}_\eta]
-n^2\frac{\beta}{2}\mathcal{E}[V^{(0)}_\eta]+n^2\frac{\beta}{2}\mathcal{E}[\eta V]\\
&&-\frac{\beta n}{2}\frac{1}{(2\pi i) }\int_0^1dt\oint_{\mathcal{L}_{2d}}
(\eta V(z)-V^{(0)}_\eta(z))u_{n,\eta}(z,t)dz,
\end{eqnarray*}
Changing the variables in the corresponding integrals, we have
\begin{eqnarray*}
\log Q_{n,\beta}[V^{(0)}_\eta]&=&\log Q_{n,\beta}^{*}+\Big(\frac{n^2\beta}{2}
+n(1-\beta/2)\Big)\log\frac{d_\eta}{2},\\
\frac{n^2\beta}{2}\mathcal{E}[V^{(0)}_\eta]&=&-\frac{3n^2\beta}{8}+\frac{n^2\beta}{2}\log\frac{d_\eta}{2}.
\end{eqnarray*}
Then   (\ref{log}) follows.
$\square$

\section{Matrix kernels for orthogonal and symplectic ensembles}\label{s:3}

The orthonormal system $\{\psi^{(n)}_{k}\}_{k=0}^\infty$ satisfies  the recursion relations
\begin{equation}\label{rec}
    \lambda\psi^{(n)}_{k}(\lambda)=a^{(n)}_{k+1}\psi^{(n)}_{k+1}(\lambda)+b^{(n)}_{k}\psi^{(n)}_{k}(\lambda)+
a^{(n)}_{k}\psi^{(n)}_{k-1}(\lambda),
\end{equation}
which define a semi-infinite Jacobi matrix $J^{(n)}$. It is known (see, e.g. \cite{PS:97}) that
\begin{equation}\label{b_J}
    |a^{(n)}_{k}|\le C,\quad |b^{(n)}_{k}|\le C,\quad |n-k|\le \varepsilon n.
\end{equation}
By orthogonality and the spectral theorem, we see
\begin{eqnarray}\label{D_j,k}
&&(D_\infty^{(n)})_{j,k}=n\,\mathrm{sign}(j-k)V'(J^{(n)})_{jk} \Rightarrow\\
\Rightarrow &&(D_\infty^{(n)})_{j,k}=0,\;|j-k|\ge 2m,\quad |(D_\infty^{(n)})_{j,k}|\le
nC,\;\;|j-n|\le nc.
\notag \end{eqnarray}
We are going to use the formula for $S_{n,\beta}$, obtained in \cite{Wi:99} (see also
\cite{DGKV:07}).
In order to present this formula, introduce some more notations:
\begin{align*}
     \Phi_1^{(n)} &:= (\psi_{n-2m+1}^{(n)} , \psi_{n-2m+2}^{(n)},\ldots ,\psi_{n-1}^{(n)})^T,\\
     \Phi_2^{(n)} &:= (\psi_n^{(n)},\psi_{n+1}^{(n)},\ldots , \psi_{n+2m-2}^{(n)})^T,
\end{align*}
and
\begin{equation*}
     M_{rs} := (\epsilon \Phi_r^{(n)}, (\Phi_s^{(n)})^T),\qquad D_{rs}:=((\Phi_r^{(n)})',
     (\Phi_s^{(n)})^T),\quad 1\le r,s\le 2.
\end{equation*}
Observe that $M_\infty^{(n)}D_\infty^{(n)}=1$ together with $(D_\infty^{(n)})_{jk}=0$
for $|j-k|\ge 2m$ implies
\begin{equation*}
     T_n=1-D_{12}M_{21}.
\end{equation*}
 Then have (see \cite{DGKV:07})
\begin{align}\notag
     S_{n,1}(\lambda,\mu) &= K_{n,2}(\lambda,\mu)+\Phi_1(\lambda)^TD_{12}
     \epsilon \Phi_2(\mu)-\Phi_1(\lambda)^T\hat G\epsilon \Phi_1(\mu),\\
     \hat G &:=D_{12}M_{22} (1-D_{21}M_{12})^{-1}D_{21}\label{r_S}\\
     S_{n/2,4}(\lambda,\mu)&= K_{n,2} (\lambda,\mu) +\Phi_2(\lambda)^TD_{12}
     \epsilon \Phi_1(\mu)-\Phi_2(\lambda)^TG\epsilon \Phi_2(\mu),\notag\\
     G&:= -D_{21}(1-M_{12}D_{21})^{-1}M_{11}D_{12},\notag
\end{align}
where $K_{n,2}$ is defined in (\ref{K_2}).

\medskip

\noindent\textit{Proof of Theorem \ref{t:1}.}
Using (\ref{r_S}),  it is straightforward to see that
the first assertion of Theorem
\ref{t:1} follows from the following lemma.

\begin{lemma}\label{l:2}
Given any smooth functions $f,g$ and any fixed $A>0$
there exists a $C>0$ such that for all $n\ge 2m$ and all
$j,k\in \{n-A,\ldots ,n+A\}$ one has
\begin{eqnarray}\label{l2.1}
&(i)&  \bigg|\int\epsilon(
f\psi_j^{(n)})(\lambda)g(\lambda)\psi_k^{(n)})(\lambda)d\lambda\bigg|\le\frac{C}{n}
\Big(||f||_\infty+||f'||_\infty\Big)\Big(||g||_\infty+||g'||_\infty\Big);\\
&(ii)&\epsilon( f\psi_j^{(n)})(\lambda)|\le\frac{C}{\sqrt{n}}\Big(||f||_\infty+||f'||_\infty\Big);\quad
(iii)\;\; \log \det (T_n)\ge C.
\notag\end{eqnarray}
\end{lemma}
Indeed, taking in (i) $f=g=1$, we obtain  that all entries of $M^{(n)}_\infty$ which are used in
(\ref{r_S}) are bounded by $Cn^{-1}$, hence, having that $|\det T|^{-1}\le C$ we obtain that $G$ and $\hat G$
have entries bounded by $nC$. This proves representation (\ref{Sn.1}).

\textit{Proof of Lemma \ref{l:2}}.
Assertions (i) and (ii) of Lemma \ref{l:2} will be derived from the asymptotics of the orthogonal
polynomials in Section \ref{s:ap1}.
We now prove assertion (iii), using Theorem \ref{t:2}.

Note  that without lost of generality we can assume that $\sigma\subset (-1,1)$ and $v^*=0$ in
(\ref{cond_rho}).

Similarly to the proof of Theorem \ref{t:2} we choose $\varepsilon$ in (\ref{s_e}) small enough to
have all zeros of $P$ of (\ref{P}) outside
of $\sigma_\varepsilon$ and use the results of \cite{PS:07} that if we replace in all definition
(\ref{p(la)}) -- (\ref{p_nl}) the integration in $\mathbb{R}$ by integration with respect to
$\sigma_\varepsilon$, then the new $Q_{n,\beta}^{(\varepsilon)}[V]$ will differ from $Q_{n,\beta}[V]$
by the factor $(1+O(e^{-nc}))$, where $c>0$ does not depend on $n$, but depends on $\varepsilon$.
Moreover, the new marginal densities will differ  from (\ref{p_nl}) by an additive error
$O(e^{-nc})$. Hence  starting from now, we assume that this replacement is made and all integrals below are
in $\sigma_\varepsilon$.

Consider the "approximating" function $H_a$ (Hamiltonian) (cf (\ref{p(la)}))
\begin{eqnarray}
H_a(\lambda_1\dots \lambda_n)&=&-n\sum  V^{(a)}(\lambda_i)+
\sum_{i\not=j}\log|\lambda_i-\lambda_j|\Big(\sum_{\alpha=1}^q\mathbf{1}_{\sigma_{\alpha,\varepsilon}}(\lambda_i)
\mathbf{1}_{\sigma_{\alpha,\varepsilon}}(\lambda_j)\Big)-n^2\Sigma^*,\notag\\
V^{(a)}(\lambda)&=&\sum_{\alpha=1}^qV^{(a)}_\alpha(\lambda),
\label{H_a}\end{eqnarray}
where $V^{(a)}_\alpha(\lambda)$ is defined in (\ref{V^a}), and $\Sigma^*$ is defined in (\ref{S*}).
Then
\begin{eqnarray}
H(\lambda_1\dots \lambda_n)&=&H_a(\lambda_1\dots \lambda_n)+\Delta H(\lambda_1\dots \lambda_n),
\quad \lambda_1,\dots,\lambda_n\in\sigma_\varepsilon,\label{DeltaH}\\
\Delta H(\lambda_1\dots \lambda_n)&=&\sum_{i\not=j}\log|\lambda_i-\lambda_j|\sum_{\alpha\not=\alpha'}
\mathbf{1}_{\sigma_{\alpha,\varepsilon}}(\lambda_i)\mathbf{1}_{\sigma_{\alpha',\varepsilon}}(\lambda_j)-2n\sum_{j=1}^n\widetilde V(\lambda_j)+n^2\Sigma^*,\notag\\
\widetilde V(\lambda)&=&\sum_{\alpha=1}^q\mathbf{1}_{\sigma_{\alpha,\varepsilon}}(\lambda)
\int_{\sigma\setminus\sigma_\alpha}\log|\lambda-\mu|\rho(\mu)d\mu.
\notag\end{eqnarray}
Set
\begin{equation}\label{Q^a}
Q^{(a)}_{n,\beta}=\int_{\sigma_\varepsilon^n}e^{\beta H_a(\lambda_1\dots \lambda_n)}d\lambda_1\dots
d\lambda_n.
\end{equation}
 By the Jensen inequality, we have
\begin{equation}\label{Bog}
\beta\left\langle\Delta H\right\rangle_{H^a,\beta}
\le\log Q_{n,\beta}[V]-\log Q^{(a)}_{n,\beta}\le\beta\left\langle\Delta H\right\rangle_{H,\beta},
\end{equation}
where
\[<\dots>_{H,\beta}:=Q_{n,\beta}^{-1}\int(\dots)e^{\beta H(\bar\lambda)}d\bar\lambda,\quad
<\dots>_{H_a,\beta}:=(Q_{n,\beta}^{(a)})^{-1}\int(\dots)e^{\beta H_a(\bar\lambda)}d\bar\lambda.\]
Let us estimate the r.h.s. of (\ref{Bog}) for $\beta=2$.
\begin{eqnarray*}
\left\langle\Delta H\right\rangle_{H,\beta}=n(n-1)\int p_{2,\beta}^{(n)}(\lambda,\mu)
\log|\lambda-\mu|\sum_{\alpha\not=\alpha'}^q
\mathbf{1}_{\sigma_{\alpha,\varepsilon}}(\lambda)\mathbf{1}_{\sigma_{\alpha',\varepsilon}}(\mu)
-2 n^2\big(\widetilde V,p_{1,\beta}^{(n)}\big)+n^2L^*.
\end{eqnarray*}
Here and below $(.,.)$ means the inner product in $L_2(\sigma_\varepsilon)$.
But using the definition of $\widetilde V$  and $L^*$ we can  rewrite the r.h.s. above as
\begin{eqnarray}\label{t1.2}
&&\left\langle\Delta H\right\rangle_{H,\beta}=\sum_{\alpha\not=\alpha'}
\int \mathbf{1}_{\sigma_{\alpha,\varepsilon}}(\lambda)
\mathbf{1}_{\sigma_{\alpha',\varepsilon}}(\mu)\log|\lambda-\mu|k^{(n)}_{\beta}(\lambda,\mu)d\lambda d\mu\\
&&+ n^2\sum_{\alpha\not=\alpha'}
L\Big[\mathbf{1}_{\sigma_{\alpha,\varepsilon}}\big(p_{1,\beta}^{(n)}-\rho\big),
\mathbf{1}_{\sigma_{\alpha',\varepsilon}}
\big(p_{1,\beta}^{(n)}-\rho\big)\Big]\le
C,\notag
\end{eqnarray}
where the kernel $k^{(n)}_{\beta}$ is defined in (\ref{k_n}) with $\eta=1$ and $h=0$.
The term with $\delta(\lambda-\mu)$ from (\ref{k_n}) gives zero contribution here, because
in our integration domain $|\lambda-\mu|\ge \delta$ (see (\ref{s_e})).
The last inequality in (\ref{t1.2}) is obtained as follows.
Since
$\mathrm{dist\,}\{\sigma_{\alpha,\varepsilon},\sigma_{\alpha',\varepsilon}\}\ge\delta$, and
$\sigma_\varepsilon\subset[-1,1]$,  we can
construct 6-periodic even function
$\widetilde L_{\alpha,\alpha'}(\lambda)$ with 12 derivatives  such that
\begin{equation}\label{L_a,a'}
\widetilde L_{\alpha,\alpha'}(\lambda-\mu)=\log|\lambda-\mu|,\; \lambda\in\sigma_{\alpha,\varepsilon},
\,\mu\in\sigma_{\alpha',\varepsilon},\quad \widetilde L_{\alpha,\alpha'}(\lambda)=0,\;
|\lambda|>5/2.
\end{equation}
  Then,
\begin{equation}\label{exp_L}
L_{\alpha,\alpha'}(\lambda-\mu)=\sum_{k=-\infty}^{\infty}c_ke^{ik\pi(\lambda-\mu)/3},\quad
\lambda\in\sigma_{\alpha,\varepsilon},
\,\mu\in\sigma_{\alpha',\varepsilon},\quad |c_k|\le Ck^{-12}.
\end{equation}
Note that we  need 12 derivatives
to have  for the Fourier coefficients of $L_{\alpha,\alpha'}$ the above bound, which will be used
later. Hence
\begin{eqnarray*}&&\int \mathbf{1}_{\sigma_{\alpha,\varepsilon}}(\lambda)\mathbf{1}_{\sigma_{\alpha',\varepsilon}}(\mu)
\log|\lambda-\mu|k^{(n)}_{\beta}(\lambda,\mu)d\lambda d\mu
\\&&=
\sum_{k=-\infty}^{\infty}c_k\int \mathbf{1}_{\sigma_{\alpha,\varepsilon}}(\lambda)
\mathbf{1}_{\sigma_{\alpha',\varepsilon}}(\mu)e^{ik\pi\lambda/3}e^{-ik\pi\mu/3}
k^{(n)}_{\beta}(\lambda,\mu)d\lambda d\mu
=O(1).
\end{eqnarray*}
Here we used the well known bound for $\beta=2$ (see e.g. \cite{PS:07})
\begin{equation}\label{PS07}
\int \varphi(\lambda)\overline{\varphi(\mu)}
k^{(n)}_{\beta}(\lambda,\mu)d\lambda d\mu=
\int |\varphi(\lambda)-\varphi(\mu)|^2K_{n,2}^2(\lambda,\mu)d\lambda d\mu\le
C||\varphi'(\lambda)||_\infty.
\end{equation}
For each term of the second sum in (\ref{t2.1}) we have similarly
\begin{eqnarray}\label{exp_L1}
&&L\big[\mathbf{1}_{\sigma_{\alpha,\varepsilon}}(p_{1,\beta}^{(n)}-\rho),
\mathbf{1}_{\sigma_{\alpha',\varepsilon}}(p_{1,\beta}^{(n)}-\rho)\big]
\\&&=
\sum_{k=-\infty}^{\infty}c_k\left(\mathbf{1}_{\sigma_{\alpha,\varepsilon}}(p_{1,\beta}^{(n)}-\rho),
e^{ik\pi\lambda/2}\right)
\left(\mathbf{1}_{\sigma_{\alpha',\varepsilon}}(p_{1,\beta}^{(n)}-\rho),
e^{-ik\pi\lambda/2}\right)=O(n^{-2}).
\notag\end{eqnarray}
This follows from the results
of \cite{DKMVZ:99}, according to which for any smooth  $\varphi$
\begin{equation}\label{DKMVZ.1}
   \int\varphi(\lambda)\Big(p_{1,2}^{(n)}(\lambda)-\rho(\lambda)\Big)d\lambda=O(n^{-1})||\varphi'||_\infty.
\end{equation}

To estimate the l.h.s. of (\ref{Bog}), we  study first the structure of $Q^{(a)}_{n,\beta}$.
Since $H_a$ does not contain terms $\log|\lambda_i-\lambda_j|$ with
$\lambda_i\in\sigma_{\alpha,\varepsilon}, \lambda_j\in\sigma_{\alpha',\varepsilon}$ with
$\alpha\not=\alpha'$,  it can be written as
\begin{eqnarray}\label{Q_k}\notag
Q^{(a)}_{n,\beta}&=&\sum_{ k_1+\dots+ k_q=n}\frac{n!}{k_1!\dots k_q!}Q_{\bar k,\beta}[V^{(a)}],\\
\quad Q_{\bar k,\beta}&=&\int_{\sigma_\varepsilon^n}
\prod_{j=1}^{k_1}\mathbf{1}_{\sigma_{1,\varepsilon}}(\lambda_j)\dots
\prod_{j=k_1+\dots+k_{q-1}+1}^{n}\mathbf{1}_{\sigma_{q,\varepsilon}}(\lambda_j)e^{\beta
H_a(\lambda_1\dots \lambda_n)}d\lambda_1\dots d\lambda_n\\
&&=e^{-n^2\beta\Sigma^*/2}\prod_{\alpha=1}^qQ_{k_\alpha,\beta}[nV_a^{(\alpha)}/k_\alpha],\notag
\end{eqnarray}
where the "effective potentials" $\{V_a^{(\alpha)}\}_{\alpha=1}^q$ are defined in (\ref{V^a}).
Take $\{k_\alpha^*\}_{\alpha=1}^q$ of
(\ref{mu^*}) and set
\begin{equation}\label{kappa}
 \kappa_{\bar k}:=\frac{k_1^*!\dots k_q^*!}{k_1!\dots k_q!}Q_{\bar
k;\beta}/Q_{\bar k^*;\beta}.
\end{equation}
It is evident that
\begin{eqnarray}\label{Q_a.1}\notag
Q^{(a)}_{n,\beta}&=&\frac{n!}{k_1^*!\dots k_q^*!}Q_{\bar k^*;\beta}\sum_{ k_1+\dots+ k_q=n}\kappa_{\bar
k}.
\end{eqnarray}
Choose $\widetilde\varepsilon$ sufficiently small to provide that
$|\mu^*_\alpha n/k_\alpha-1|\le\varepsilon_1^{(\alpha)}$
for $\alpha=1,\dots,q$, if $|\bar k-\bar k^*|\le\widetilde\varepsilon n$,
where $\varepsilon_1^{(\alpha)}$ is chosen
for $\sigma_{\alpha,\varepsilon}$ and $(n/k_\alpha )V_a^{(\alpha)}$, according to Theorem \ref{t:2}.

Let us prove  that there exist $n,\bar k$-independent $C_*,c_*>0$ and $\bar a\in\mathbb{R}^q$ such
that
\begin{equation}\label{b_kappa}
\kappa_{\bar k}\le \left\{\begin{array}{ll}C_*e^{-c_*(\bar k-\bar k^*,\bar k-\bar k^*)+(\bar a,\bar k-\bar k^*)},
&|\bar k-\bar k^*|\le\widetilde\varepsilon n,\\ C_*e^{-c_0n^2}&|\bar k-\bar k^*|\ge\widetilde\varepsilon
n.\end{array}\right.
\end{equation}
The proof of the first bound is based on Theorem \ref{t:2}.
We write
\begin{eqnarray}\label{t1.3}
\log\Big(k^*_\alpha!Q_{k_\alpha,\beta}/k_\alpha!Q_{k^*_\alpha,\beta}\Big)&=&
\log\Big(Q_{k_\alpha;\beta}/Q_{k_\alpha;\beta}^{*}\Big)
-\log\Big(Q_{k^*_\alpha,\beta}/Q_{k^*_\alpha,\beta}^{*}\Big)\\
&&+
\log\Big(k^*_\alpha!Q_{k_\alpha,\beta}^{*}/k_\alpha!Q_{k^*_\alpha,\beta}^{*}\Big),\notag
\end{eqnarray}
where $Q_{k_\alpha;\beta}^{*}$ is defined in (\ref{Sielb}). Then (\ref{log}) yields
\begin{eqnarray}\label{t1.4}
&&\log\Big(Q_{k_\alpha,\beta}/Q_{k_\alpha,\beta}^{*}\Big)=\frac{k_\alpha^2\beta}{2}
\Big(\mathcal{E}_{n/k_\alpha}+\frac{3}{4}\Big)
- k_\alpha I(k_\alpha)+O(1),\\
&&\notag I(k_\alpha):=\frac{\beta }{4\pi i}\int_0^1dt
\oint\Big(\frac{n}{k_\alpha}V_a(\zeta)-V^{(0)}_{n/k_\alpha}(\zeta)\Big)u_{n/k_\alpha}(\zeta,t)d\zeta
+\Big(1-\beta/2\Big)\log (d_{n/k_\alpha}/2),
\end{eqnarray}
where $\mathcal{E}_{n/k_\alpha}=\mathcal{E}[(n/k_\alpha )V^{(a)}_\alpha]$ is the equilibrium energy (\ref{E_V}),
$u_{n/k_\alpha}(., t)$, $V^{(0)}_{n/k_\alpha}$, and $d_{n/k_\alpha}$ are defined in (\ref{log}).
Using (\ref{t2.2}) and (\ref{t2.2*}), we get for $I(k_\alpha)$ of (\ref{t1.4})
\begin{eqnarray}\label{dif_I}
 k_\alpha I(k_\alpha)- k_\alpha^* I(k_\alpha^*)
=b^{(\alpha)}(k_\alpha-k^*_\alpha)+O(1)
+O((k_\alpha-k^*_\alpha)^2/n),
\end{eqnarray}
where the concrete form of $b^{(\alpha)}$ is not important for us.
Moreover,  formula (\ref{Sielb}) for $n=k_\alpha=k_\alpha^*+l$ and $n=k_\alpha^*$  yields
\begin{eqnarray}\notag
&&\hskip-2cm\log\Big(k_\alpha^*!Q_{k_\alpha,\beta}^{*}/k_\alpha!Q_{k^*_\alpha,\beta}^{*}\Big)
-l\log \Big(\sqrt{2\pi}/\Gamma(\beta/2)\Big)\\\notag&&
=-\Big(\frac{\beta}{4}(k^*_\alpha +l)^2
+\frac{k^*_\alpha +l}{2}(1-\beta/2)\Big)\log\frac{\beta(k^*_\alpha +l)}{2}\\
&&+\Big(\frac{\beta}{4}(k^*_\alpha )^2
+\frac{k^*_\alpha }{2}(1-\beta/2)\Big)\log\frac{\beta k^*_\alpha}{2}+\sum_{j=1}^l\log\Gamma(\beta(k^*_\alpha +j)/2)
\notag\\&&=
l(1-\beta/2)\Big(\log(\beta k^*_\alpha/2)+\frac{1}{2}\Big)-\frac{3}{4}\beta k^*_\alpha l
-\frac{3}{8}\beta  l^2+O(l^3/k^*_\alpha).
\label{t1.5}\end{eqnarray}
Here we  used  the Stirling formula
\begin{multline*}
\log\Gamma(\beta(k^*_\alpha +j)/2)=\Big(\beta(k^*_\alpha +j)/2-1/2\Big)\log(\beta(k^*_\alpha +j)/2)-
\beta(k^*_\alpha +j)/2+O(n^{-1})\\
=\Big(\beta(k^*_\alpha +j)/2-1/2\Big)\log(\beta k^*_\alpha /2)+\Big(\beta(k^*_\alpha
+j)/2-1/2\Big)\frac{j}{ k^*_\alpha}-\beta(k^*_\alpha +j)/2+O(l/k_\alpha)
\end{multline*}
and the representation
\[\log\frac{\beta(k^*_\alpha +l)}{2}=\log\frac{\beta k^*_\alpha }{2}+\frac{l}{k^*_\alpha}-
\frac{l^2}{2(k^*_\alpha)^2}+O(l^3/k_\alpha^3).\]
Moreover,
\[\frac{3k_\alpha^2\beta}{8}-
\frac{(3k^*_\alpha)^2\beta}{8}=\frac{3\beta l k_\alpha}{4}+
\frac{3\beta l^2}{8}.\]
The last relation and (\ref{t1.3}) -- (\ref{t1.5}) yield
\begin{eqnarray}\label{t1.6}
&&\hskip-1.5cm\log\Big(k^*_\alpha!Q_{k_\alpha,\beta}/k_\alpha!Q_{k^*_\alpha,\beta}\Big)
=\frac{k_\alpha^2\beta}{2}\mathcal{E}_{n/k_\alpha}-
\frac{(k_\alpha^*)^2\beta}{2}\mathcal{E}_{n/k_\alpha^*}
+(k_\alpha-k^*_\alpha) a^{(\alpha)}\\&&+(k_\alpha-k^*_\alpha)(1-\beta/2)\log n
+O(1),
\notag\end{eqnarray}
where
\[a^{(\alpha)} =(1-\beta/2)\Big(\log(\beta k^*_\alpha/2n)+\frac{1}{2}\Big)+\log \Big(\sqrt{2\pi}/\Gamma(\beta/2)\Big)
+b^{(\alpha)}\]
with $b^{(\alpha)}$ of (\ref{dif_I}).
To estimate the difference of the energies we
introduce the densities
\begin{equation}\label{t1.7}
   \widetilde\rho_{ n/k_\alpha}^{(\alpha)}(\lambda)=\frac{k_\alpha}{n}\mathbf{1}_{\sigma_{\alpha,\varepsilon}}(\lambda)
   \rho_{n/k_\alpha}(\lambda),\quad
    \rho^{(\alpha)}(\lambda)=\rho(\lambda)\mathbf{1}_{\sigma_{\alpha}}(\lambda).
\end{equation}
Then
\begin{eqnarray*}
k_\alpha^2\mathcal{E}_{n/k_\alpha}&=&n^2\Big(L\Big[
\widetilde\rho_{ n/k_\alpha}^{(\alpha)},\widetilde\rho_{ n/k_\alpha}^{(\alpha)}\Big]
-\big( V_a,\widetilde\rho_{ n/k_\alpha}^{(\alpha)}\big)\Big)
\\
&&
=n^2\Big(L\Big[
\widetilde\rho_{ n/k_\alpha}^{(\alpha)},\widetilde\rho_{ n/k_\alpha}^{(\alpha)}\Big]
-\big( V,\widetilde\rho_{ n/k_\alpha}^{(\alpha)}\big)
+ 2L\Big[
\widetilde\rho_{ n/k_\alpha}^{(\alpha)},\rho-\rho^{(\alpha)}\Big]\Big).
\notag\end{eqnarray*}
Hence   taking
\begin{eqnarray*}n^2\mathcal{E}^{(\alpha)}:=n^2\Big(L\big[
\rho^{(\alpha)},\rho^{(\alpha)}\big]
-\left(V,\rho^{(\alpha)}\right)
+ 2L\big[
\rho^{(\alpha)},\rho-\rho^{(\alpha)}\big]\Big),
\end{eqnarray*}
we obtain
\begin{eqnarray}\notag
k_\alpha^2\mathcal{E}_{n/k_\alpha}-n^2\mathcal{E}^{(\alpha)}
&=&n^2\Big(L\Big[\widetilde\rho_{ n/k_\alpha}^{(\alpha)}-\rho^{(\alpha)},
\widetilde\rho_{ n/k_\alpha}^{(\alpha)}-\rho^{(\alpha)}\Big] +
2L\Big[\widetilde\rho_{ n/k_\alpha}^{(\alpha)}-\rho^{(\alpha)},\rho^{(\alpha)}\Big]
\\\label{t1.8}
&&+2L\Big[\widetilde\rho_{ n/k_\alpha}^{(\alpha)}-\rho^{(\alpha)},\rho-\rho^{(\alpha)}\Big]
-\big(V,\widetilde\rho_{n/k_\alpha}^{(\alpha)}-\rho^{(\alpha)}\big)\\
&\le&n^2L\Big[\widetilde\rho_{ n/k_\alpha}^{(\alpha)}-\rho^{(\alpha)},
\widetilde\rho_{ n/k_\alpha}^{(\alpha)}-\rho^{(\alpha)}\Big]=:n^2L_{ n/k_\alpha}.
\notag\end{eqnarray}
Here  we used  that in view of (\ref{cond_rho}) and our assumption  $v^*=0$, we have
\begin{eqnarray*}
&&v(\lambda)=\int\log |\lambda-\mu|\rho(\mu) d\mu-V(\lambda)=0,\,
\lambda\in\sigma_\alpha,\quad\quad
v(\lambda)<0,\,\lambda\not\in\sigma_\alpha,
\\
&&\widetilde\rho_{ n/k_\alpha}^{(\alpha)}(\lambda)-\rho^{(\alpha)}(\lambda)=
\widetilde\rho_{ n/k_\alpha}^{(\alpha)}(\lambda)\ge0,
\quad \lambda\not\in\sigma_\alpha
\notag\end{eqnarray*}
which implies
\begin{equation}\label{t1.8c}
L\Big[\widetilde\rho_{ n/k_\alpha}^{(\alpha)}-\rho^{(\alpha)},\rho\Big]
-\big(V,\widetilde\rho_{n/k_\alpha}^{(\alpha)}-\rho^{(\alpha)}\big)=
\big(v,\widetilde\rho_{n/k_\alpha}^{(\alpha)}-\rho^{(\alpha)}\big)\le 0.\end{equation}
Note that for any function
$\varphi:\mathrm{supp\,}\varphi\subset[-1,1],\int\varphi=\varphi^*$
\begin{eqnarray}\label{t1.8a}
L\Big[\varphi,\varphi\Big]=L\Big[\varphi-\varphi^*\psi_*,\varphi-\varphi^*\psi_*\Big]
-|\varphi^*|^2\log 2\le-|\varphi^*|^2\log 2,
\end{eqnarray}
where $\psi_*(\lambda)=\pi^{-1}(1-\lambda^2)^{-1/2}\mathbf{1}_{[-1,1]}$, and we used the well known properties of
$\psi_*$:
\[\int_{-1}^{1}\log|\lambda-\mu|\psi_*(\mu)d\mu=-\log2,\, \lambda\in[-1,1],\quad\quad
\int_{-1}^{1}\psi_*(\mu)d\mu=1.\]
Since, by (\ref{t1.7}),
\[\int\rho_{ n/k_\alpha}^{(\alpha)}(\lambda)d\lambda=k_\alpha/n,\quad\int\rho^{(\alpha)}(\lambda)d\lambda=\mu^*_\alpha,
\]
we get
\begin{equation}\label{t1.8b}
n^2L_{ n/k_\alpha}\le
-\log2|k_\alpha/n-\mu^*_\alpha|^2n^2.
\end{equation}
In view of (\ref{t1.6}) and (\ref{t1.8}), to obtain the estimate for the difference of energies,
it suffices to obtain  the bound for
$(k_\alpha^*)^2\mathcal{E}_{n/k^*_\alpha}-n^2\mathcal{E}^{(\alpha)}$ from below.
But it follows from (\ref{t2.2*})
and (\ref{rho}) that
  \begin{equation}\label{t1.8d}
\widetilde\rho_{n/k_\alpha^*}^{(\alpha)}(\lambda)
-\rho^{(\alpha)}(\lambda)=O(|d^*_\alpha|^{1/2}),\quad
v(\lambda)=O(|d^*_\alpha|^{3/2}),\quad\lambda\not\in\sigma_\alpha,
\end{equation}
where
\[d^*_\alpha:=k^*_\alpha/n-\mu^*_\alpha.\]
 Since
$|d^*_\alpha|\le 2n^{-1}$, the l.h.s. of (\ref{t1.8c})
is $O(|d^*_\alpha|^{3})=O(n^{-3})$ for $k_\alpha=k_\alpha^*$.
 Hence
 \[(k_\alpha^*)^2\mathcal{E}_{n/k^*_\alpha}-n^2\mathcal{E}^{(\alpha)}=n^2L_{
 n/k_\alpha^*}+O(n^{-1}).\]
In order to estimate $L_{ n/k_\alpha^*}$, observe that if we denote $v_{n/k^*_\alpha}(\lambda)$ the
l.h.s. of (\ref{cond_rho}) for $(n/k^*_\alpha) V$, then we have
\[v_{n/k^*_\alpha}(\lambda)-v(\lambda)=d^*_\alpha V^{(a)}_\alpha(\lambda)+v^*_{n/k^*_\alpha},\quad\lambda\in\sigma_{\alpha}
\cap\mathrm{supp\,}\widetilde\rho_{n/k_\alpha^*}^{(\alpha)}.\]
where
\[v^*_{n/k^*_\alpha}=\sup v_{n/k^*_\alpha}(\lambda)=v_{n/k^*_\alpha}(c_\alpha)=O(d^*_\alpha),\]
and $c_\alpha$ is a middle point of $\mathrm{supp\,}\widetilde\rho_{n/k_\alpha^*}^{(\alpha)}$.
Moreover, the above argument implies that the contribution of the integrals
over the domain $\sigma_{\alpha,\varepsilon}\setminus (\sigma_{\alpha}
\cap\mathrm{supp\,}\widetilde\rho_{n/k_\alpha^*}^{(\alpha)})$ is $O(|d^*_\alpha|^{5/2}) $. Thus
\begin{eqnarray}\label{b_E}
n^2L_{ n/k_\alpha^*}=O(1),&\mathrm{and}&k_\alpha^2\mathcal{E}_{n/k_\alpha}-(k_\alpha^*)^2\mathcal{E}_{n/k_\alpha^*}
\le -\log 2(k_\alpha-k^*_\alpha)^2+O(1).
\end{eqnarray}
 Then
we obtain the first bound of (\ref{b_kappa}) from (\ref{t1.6}) and the last inequality,
if  take the sum with respect to
$\alpha=1,\dots,q$ and take into account that $\sum_\alpha(k_\alpha-k^*_\alpha)=0$.

To obtain the second bound of (\ref{b_kappa}), we use  the inequality
\[
\log Q_{k_\alpha,\beta}\le \frac{\beta k_\alpha^2}{2}\mathcal{E}_{n/k_\alpha}+Ck_\alpha\log k_\alpha
\]
with some universal $C$ (see \cite{BPS:95}).
Using the inequality for $k_\alpha$ and $k_\alpha^*$, we get in view of (\ref{b_E})
\begin{eqnarray*}
\log \Big(Q_{k_\alpha,\beta}/Q_{k_\alpha^*,\beta}\Big)&\le &
-\frac{\beta}{2}\log 2|k_\alpha-k_\alpha^*|^2+O(n\log n).
\end{eqnarray*}
Hence
we obtain the second inequality of (\ref{b_kappa}).
Now we are ready to find a bound for the l.h.s. of (\ref{Bog}). We have
\begin{eqnarray}\label{t1.9}
\left\langle\Delta H\right\rangle_{H^a,\beta}&=&\sum_{ k_1+\dots+ k_q=n}\kappa_{\bar k}
\left\langle\Delta H\right\rangle_{\bar k}\Big(\sum_{k_1+\dots+ k_q=n}\kappa_{\bar
k}\Big)^{-1},\\
\left\langle\Delta H\right\rangle_{\bar k}&:=&\sum_{\alpha\not=\alpha'}k_\alpha k_{\alpha'}
L\Big[p^{(k_\alpha)}_{1,\beta},p^{(k_{\alpha'})}_{1,\beta}\Big]
-2n\sum_{\alpha}k_\alpha\left(\widetilde V(\lambda),p^{(k_\alpha)}_{1,\beta}\right)
\notag\\
&&+n^2\sum_{\alpha\not=\alpha'}
L\Big[\rho^{(\alpha)},\rho^{(\alpha')}\Big]\notag\\
&=&\sum_{\alpha\not=\alpha'}k_\alpha k_{\alpha'}
L\Big[p^{(k_\alpha)}_{1,\beta}-\frac{n}{k_\alpha}\rho^{(\alpha)},
p^{(k_{\alpha'})}_{1,\beta}-\frac{n}{k_{\alpha'}}\rho^{(\alpha')}\Big].\notag\end{eqnarray}
Then, for $|\bar k-\bar k^*|\le\widetilde\varepsilon n$ we write similarly to (\ref{exp_L}) and (\ref{exp_L1})
\begin{eqnarray*}
&& \bigg|L\Big[p^{(k_\alpha)}_{1,\beta}-\frac{n}{k_\alpha}\rho^{(\alpha)},
p^{(k_{\alpha'})}_{1,\beta}-\frac{n}{k_{\alpha'}}\rho^{(\alpha')}\Big]\bigg|\\&&=
\bigg|\sum_{j=-\infty}^\infty c_j\big(p^{(k_\alpha)}_{1,\beta}-\frac{n}{k_\alpha}\rho^{(\alpha)},e^{ij\pi\lambda/3}\big)
\big(p^{(k_{\alpha'})}_{1,\beta}-\frac{n}{k_{\alpha'}}\rho^{(\alpha')},e^{ij\pi\lambda/3}\big)\bigg|\\
&&\le\sum_{j=-\infty}^\infty
|c_j|\bigg(\Big|\big(p^{(k_\alpha)}_{1,\beta}-\rho_{n/k_\alpha},e^{ij\pi\lambda/3}\big)\Big|^2+
\Big|\big(p^{(k_{\alpha'})}_{1,\beta}-\rho_{n/k_{\alpha'}},e^{ij\pi\lambda/3}\big)\Big|^2\\
&&+\Big|\big(\rho_{n/k_\alpha}-\frac{n}{k_\alpha}\rho^{(\alpha)},e^{ij\pi\lambda/3}\big)\Big|^2
+\Big|\big(\rho_{n/k_{\alpha'}}-\frac{n}{k_{\alpha'}}\rho^{(\alpha')},e^{ij\pi\lambda/3}\big)\Big|^2\bigg)\\
&&\le O(n^{-2})+
C\big(n/k_\alpha-(\mu_\alpha^*)^{-1}\big)^2+\big(n/k_{\alpha'}-(\mu_{\alpha'}^*)^{-1}\big)^2.
\end{eqnarray*}
To get the  bound $O(n^{-2})$ for the first two terms in the r.h.s. here, we  used   (\ref{t2.1a}) for
$\varphi_k=e^{ik\pi\lambda/3}$ and the bound for $c_j$ from (\ref{exp_L}),
and for the last two terms we used (\ref{exp_L}), combined with the estimate
\begin{equation*}
    \Big|\big(\rho_{n/k_\alpha}-\frac{n}{k_\alpha}\rho^{(\alpha)},e^{ij\pi\lambda/3}\big)\Big|\le
    C|n/k_\alpha-(\mu^*_\alpha)^{-1}|.
\end{equation*}
The estimate follows from (\ref{rho}) and (\ref{t2.2*}), since these relations mean that $\rho_\eta$
is differentiable with respect to $\eta$ at $\eta=1$, and  $(\rho_\eta)'_\eta\in L_1[\sigma_{\alpha,\varepsilon}]$.
For $|\bar k-\bar k^*|\ge\widetilde\varepsilon n$,  in view of the second line of
(\ref{b_kappa}), it suffices to use that the r.h.s. of (\ref{t1.9}) is $O(n^2)$.
Thus we get finally
\[|\left\langle\Delta H\right\rangle_{H^a,\beta}|\le C .\]
The bound, (\ref{detT}), and (\ref{Bog}),  yield
\begin{equation*}
     \det (T_n)=\left(\frac{Q_{n,1}Q_{n/2,4}}{Q_{n,2}(n/2)!2^n}\right)^2
 \ge C\prod_{\alpha=1}^q\left(\frac{Q_{k^*_\alpha,1}Q_{k^*_\alpha/2,4}}
     {Q_{k^*_\alpha,2}(k^*_\alpha/2)!2^{k^*_\alpha}}\right)^2.
\end{equation*}
Hence  it is enough to prove that each multiplier in the r.h.s. is bounded from below.
Consider  the functions $V_t$ of (\ref{V_t}).
Then, as it was mentioned in  the proof of Theorem \ref{t:2} (iii),
the limiting equilibrium density $\rho_t$ has the form (\ref{rho_t} ), which corresponds to $V_t$.
Hence  $V_{n/k_\alpha^*,t}$ satisfies conditions of Theorem \ref{t:2} for any $t\in[0,1]$. Moreover,
 if we
introduce the matrix $T_n(t)$  for the potential
$V_t$ by the same way, as above, then $T_n(0)$ corresponds to  GOE or GSE.
Consider the function
\begin{equation}\label{L}
    L(t)=\log\hbox{det}\,T_{k^*_\alpha}(t).
\end{equation}
To prove that $|L(1)|\le C$ it is enough to prove that
\begin{equation}\label{cond_L}
|L(0)|\le C,\quad |L'(t)|\le C, \quad t\in[0,1].
\end{equation}
The first inequality here follows from the results of
\cite{Tr-Wi:98}. To prove the second inequality we use (\ref{detT}) for $V$ replaced by $V_t$. Then we get
\begin{eqnarray*}
&&L'(t)=(k^*_\alpha)^2\Big(\int \Delta V(\lambda)p^{(k^*_\alpha/2)}_{1,4,t}(\lambda)d\lambda
+\int \Delta V(\lambda)p^{(k^*_\alpha)}_{1,1,t}(\lambda)d\lambda
-2\int \Delta V(\lambda)p^{(k^*_\alpha)}_{1,2,t}(\lambda)d\lambda\Big),\\
&&\Delta V(\lambda)=(n/k_\alpha^*)V^{(a)}_\alpha(\lambda)-V^{(0)}_{n/k_\alpha^*}(\lambda).
\end{eqnarray*}
 Using  (\ref{t2.1}), we obtain
 that the first and the second terms of (\ref{t2.1}) give zero contributions in
$L'(t)$, hence  $L'(t)=O(1)$. Thus we
have proved the second inequality in (\ref{cond_L}) and so assertion (iii) of Lemma \ref{l:2}.
As it was mentioned above, assertions  (i) and (iii) of Lemma \ref{l:2}, combined
with (\ref{D_j,k}), imply that
$F_{jk}^{(1)}$ and $F_{jk}^{(4)}$ of (\ref{Sn.1}) are bounded uniformly in $n$.
\medskip

To prove the first line of (\ref{c2.1}) for $\beta=1$,
we use (\ref{Sn.1})
\begin{eqnarray*}
\int p_{1,1}^{(n)}(\lambda)f(\lambda)d\lambda
&=&n^{-1}\int S_{n,1}(\lambda,\lambda)f(\lambda)d\lambda\\
&=&\int p_{1,2}^{(n)}(\lambda)f(\lambda)d\lambda+\sum_{j,k=-(2m-1)}^{2m-1}F^{(1)}_{jk}\int
\psi^{(n)}_{n+j}(\lambda)\epsilon\psi^{(n)}_{n+k}(\lambda)f(\lambda)d\lambda.
\end{eqnarray*}
Thus
we obtain the first line of (\ref{c2.1}) from (\ref{DKMVZ.1}) and (\ref{l2.1})(i).
For $\beta=4$ the proof is the same.

The second line of (\ref{c2.1}) follows from the first one in view of Lemma \ref{l:1}.

\medskip

To obtain (\ref{fact}), observe  that we  proved already that the l.h.s. of (\ref{fact}) is
more than the r.h.s. Hence  we are left to prove the opposite inequality. To this aim we use
the inequality (\ref{Bog}) and the relation (\ref{t1.2}). Then each term in the first sum of
(\ref{t1.2}) can be estimated by the same way, as for $\beta=2$ by using the second line of (\ref{c2.1})
instead of (\ref{PS07}).
Each term in the second sum in (\ref{t1.2}) can be estimated by the same way, as for $\beta=2$,
if we use the first line of (\ref{c2.1}).
$\square$

\medskip

\textit{Proof of Theorem \ref{Tt:2}}. Convergence of the diagonal entries of $K_{n,1}$ and
$K_{n/2,4}$ to the correspondent limiting expressions follows from the first assertion of Theorem
\ref{t:1}, (ii) of (\ref{l2.1}) and convergence of $K_{n,2}$ to $K_{\infty}$. The same is valid for
12-entries of $K_{n,1}$ and $K_{n/2,4}$. Moreover,
since $\epsilon S_{n,\beta}(\lambda,\mu)=-\epsilon S_{n,\beta}(\mu,\lambda)$, one has
\begin{equation*}
     (\epsilon S_{n,1})(\lambda,\mu)=-\int_\lambda^\mu S_{n,1}(t,\mu)\, dt ,
     \qquad(\epsilon S_{n/2,4})(\lambda,\mu)=-\int_\lambda^\mu S_{n/2,4}(t,\mu)\, dt,
\end{equation*}
which implies convergence of 21-entries of $K_{n,1}$ and $K_{n/2,4}$.$\square$

 \section{Uniform bounds for  integrals with $\epsilon (f\psi_k^{(n)})$}\label{s:ap1}
 Set
 \begin{eqnarray}\label{de_n}
   && \delta_n=n^{-2/3+\kappa},\; 0<\kappa<1/3, \\
    &&\sigma_{\pm\delta_n}=\bigcup_{\alpha=1}^q\sigma_{\alpha,\pm\delta_n},\;
    \sigma_{\alpha,\pm\delta_n}=[E_{2\alpha-1}\mp\delta_n,E_{2\alpha}\pm\delta_n],\;
    (\sigma_{\alpha,-\delta_n}\subset\sigma_\alpha\subset\sigma_{\alpha,+\delta_n}).\notag
\end{eqnarray}
Then, according to  \cite{DKMVZ:99}, we have
\begin{eqnarray}
  \psi_n^{(n)}(\lambda)&=&R_0(\lambda)\cos n\pi F_n(\lambda)\;(1+O(n^{-1})),
 \quad \lambda\in\sigma_{-\delta_n},\notag\\
 \psi_{n-1}^{(n)}(\lambda)&=&R_1(\lambda)\sin n\pi F_{n-1}(\lambda)\;(1+O(n^{-1})),
 \quad\lambda\in\sigma_{-\delta_n},
 \label{as_psi}\end{eqnarray}
 where $R_0(\lambda),\, R_{1}(\lambda)$ are some smooth functions, which
may behave like $|X^{-1/4}(\lambda)|$ near each $E_\alpha$ (see (\ref{X}) for the definition of $X$).
\begin{eqnarray}\label{F_n}
F_n(\lambda)=F(\lambda)+\frac{1}{n}m_0(\lambda),\quad F(\lambda)=\int_{\lambda}^{E_{2q}}\rho(\mu)d\mu,\quad
F_{n-1}(\lambda)=F(\lambda)+\frac{1}{n}m_1(\lambda),
\end{eqnarray}
with $\rho$ of (\ref{rho}) and smooth $m_0,m_1$, such that their first derivatives
are bounded by $|X^{-1/2}(\lambda)|$. It will be important for us that
\begin{equation}\label{rel_R}
a_n^{(n)}R_0(\lambda)R_1(\lambda)\cos(\pi(m_0(\lambda)-m_1(\lambda)))=1,
\end{equation}
where $a_n^{(n)}$ is defined in (\ref{rec}). The  relation follows  from the fact (see \cite{DKMVZ:99}) that
\[\pi a_n^{(n)}n^{-1}\Big((\psi_n^{(n)}(\lambda))'\psi_{n-1}^{(n)}(\lambda)-\psi_n^{(n)}(\lambda)
 (\psi_{n-1}^{(n)}(\lambda))'\Big)=
 \rho(\lambda)+O(n^{-1})\]
For $|\lambda-E_\alpha|\le\delta_n$ we have
 \begin{eqnarray}\notag
\psi_n^{(n)}(\lambda)&=&n^{1/6}B_{11}^{(\alpha)}Ai\left( n^{2/3}
\Phi_{\alpha}\Big((-1)^\alpha(\lambda- E_\alpha)\Big)\right)(1+O(|\lambda-E_\alpha|))
\\&&+n^{-1/6}B_{12}^{(\alpha)}Ai'\left( n^{2/3}
\Phi_{\alpha}\Big((-1)^\alpha(\lambda- E_\alpha)\Big)\right)(1+O(|\lambda-E_\alpha|))+O(n^{-1}),
\notag\\
\psi_{n-1}^{(n)}(\lambda)&=&n^{1/6}B_{21}^{(\alpha)}Ai\left( n^{2/3}
\Phi_{\alpha}\Big((-1)^\alpha(\lambda- E_\alpha)\Big)\right)(1+O(|\lambda-E_\alpha|))\label{as_psi.1}
\\&&+n^{-1/6}B_{22}^{(\alpha)}Ai'\left( n^{2/3}
\Phi_{\alpha}\Big((-1)^\alpha(\lambda- E_\alpha)\Big)\right)(1+O(|\lambda-E_\alpha|))+O(n^{-1}).
\notag
\end{eqnarray}
Moreover,
\[
|\psi_n^{(n)}(\lambda)|+|\psi_{n-1}^{(n)}(\lambda)|\le e^{-nc\,\mathrm{dist}^{3/2}\{\lambda,\sigma\}},\quad
\lambda\in\mathbb{R}\setminus\sigma_{-\delta_n}.\notag\]
 Functions $\Phi_{\alpha}$ in (\ref{as_psi}) are analytic in  some neighborhood of $0$
and  such that $\Phi_{\alpha}(\lambda)= a_{\alpha}x+O(x^2)$ with some positive $a_{\alpha}$.

The proof of Lemma \ref{l:2} is based on the proposition:
\begin{proposition}\label{p:*} Under conditions of \textbf{C1-C3} for any smooth function $f$  we have
uniformly in $\sigma_{+\delta_n}$
\begin{eqnarray}\label{as_eps}
 \epsilon (f\psi_n^{(n)})(\lambda)& =&f(\lambda)R_0(\lambda)\frac{\cos
  nF_n(\lambda)}{nF'_n(\lambda)}\mathbf{1}_{\sigma_{-\delta_n}}+\chi_n(\lambda)\\
 && + \sum_{\alpha=1}^{2q}\bigg(f(E_\alpha)\frac{n^{-1/2}B^{(\alpha)}_{11}}
 {(-1)^\alpha\Phi'_\alpha\big(0\big)}
  \Psi\left( n^{2/3}
\Phi_{\alpha}\Big((-1)^\alpha(\lambda- E_\alpha)\Big)\right)
 \notag\\&&
  +O(n^{-5/6})\bigg)\mathbf{1}_{|\lambda-E_\alpha|\le\delta_n}
 +\epsilon r_n(\lambda)+O(n^{-1}),\notag\\
 \Psi(x)&:=&\int_{-\infty}^x Ai(t)dt,
\notag\end{eqnarray}
where $|\chi_n(\lambda)|\le n^{-1/2}C$ is a piecewise constant function  which is a constant
in each $\sigma_{\alpha,-\delta_n}$ and each interval $(E_\alpha-\delta_n,E_\alpha+\delta_n)$,
and the remainder  $r_n(\lambda)$ admits the bound
\begin{equation}\label{b_er}
\int_{\sigma_{+\delta_n}}|r_n(\lambda)|d\lambda\le Cn^{-1/2-3\kappa/4}.
\end{equation}
Similar representation
is valid for $\epsilon (f\psi_{n-1}^{(n)})$ if we replace $\sin$ by $\cos$,  $R_0$ by $R_{1}$,
$F_n$ by $F_{n-1}$,
and $B_{11}^{(\alpha)}$ by $B_{21}^{(\alpha)}$.
\end{proposition}
\textit{Proof.}
Let
 $\lambda\in \sigma_{\alpha,-\delta_n}$. Then, integrating by parts
 in (\ref{as_psi}), we obtain
\begin{eqnarray}\notag
\int_{E_{2\alpha-1}+\delta_n}^\lambda f(\mu)\psi^{(n)}_n(\mu)d\mu=
f(\mu)R(\mu)\frac{\sin nF_n(\mu)}{nF'(\mu)}\Bigg|_{E_{2\alpha-1}+\delta_n}^\lambda\\-
\int_{E_{2\alpha-1}+\delta_n}^\lambda \sin
nF_n(\mu)\frac{d}{d\mu}\frac{f(\mu)R(\mu)}{nF'(\mu)}d\mu.
\label{p*.1}\end{eqnarray}
Moreover, by (\ref{as_psi.1}), we have for $|\lambda-E_{2\alpha}|\le \delta_n$
\begin{eqnarray*}
\int_{E_{2\alpha}-\delta_n}^\lambda \psi_n^{(n)}(\mu)f(\mu)d\mu=n^{-1/2}B^{(2\alpha)}_{11}f(\lambda)
  \frac{\Psi\left(n^{2/3}\Phi_{2\alpha}(\lambda-E_{2\alpha})\right)
  -\Psi\left(n^{2/3}\Phi_{2\alpha}(-\delta_n)\right)}{\Phi'(\lambda-E_\alpha)}\\+
  O(n^{-5/6})
  =n^{-1/2}B^{(2\alpha)}_{11}f(E_{2\alpha})
  \frac{\Psi\left(n^{2/3}\Phi_{2\alpha}(\lambda-E_{2\alpha})\right)
  }{\Phi'(0)}+o(n^{-1/2})\mathrm{const}+
  O(n^{-5/6}).
  \end{eqnarray*}
Similar relations are valid for integrals near $E_{2\alpha-1}$. Taking into account that the
integrals over $\mathbb{R}\setminus\sigma_{+\delta_n}$ are of the order $O(e^{-nc\delta_n^{3/2}})=
O(e^{-n^{3\kappa/2}})$, we obtain (\ref{as_eps}),  writing  $\epsilon(f\psi_n^{(n)})$ as
a sum of the above integrals  and similar ones (with integration from $\lambda$ to ${E_{2\alpha}}$).
Then, for $\lambda\in\sigma_{\alpha,-\delta_n}$
\[\chi_n(\lambda)=\frac{1}{2}\int_{E_1-\delta_n}^{E_{2\alpha-1}+\delta_n} f(\mu)\psi^{(n)}_n(\mu)d\mu-
\frac{1}{2}\int_{E_{2\alpha-1}+\delta_n}^{E_{2q}+\delta_n} f(\mu)\psi^{(n)}_n(\mu)d\mu=O(n^{-1/2}),\]
and  $r_n$ is  the sum of the terms, which are under the
integrals in the r.h.s. of (\ref{p*.1}).

 Hence  we are left to prove the bound for
  $r_n$.  Using  that
  \[F'_n(\lambda)=(2\pi)^{-1} P(\lambda)X^{1/2}(\lambda)+n^{-1}m'(\lambda),\quad
  |R(\lambda)|\le C|X^{-1/4}(\lambda)|,\]
  we have
\begin{eqnarray*}\int|r_n(\mu)|d\mu &\le&
\int_{\sigma_{-\delta_n}}\left|\frac{d}{d\mu}\frac{f(\mu)R(\mu)}{nF'(\mu)}\right|d\mu+
  O(n^{-5/6})\\
&\le& C(1+||f'||_\infty) n^{-1}\delta_n^{-3/4}= C(1+||f'||_\infty) n^{-1/2-3\kappa/4}.
\end{eqnarray*}
$\square$

\textit{Proof of Lemma \ref{l:2}}.
Using recursion relations (\ref{rec}),
it is easy to get that for any $|j|\le 2m$
\[\psi^{(n)}_{n+j}(\lambda)=f_{0j}(\lambda)\psi^{(n)}_{n}(\lambda)+f_{1j}(\lambda)\psi^{(n)}_{n-1}(\lambda),\]
where $f_{0j}$ and $f_{1j}$ are polynomials of degree at most $|j|$. Note that
since  $a^{(n)}_{k}$ and $b^{(n)}_{k}$ are bounded uniformly in $n$ for $k-n=o(n)$,
$f_{0j}$ and $f_{1j}$ have  coefficients, bounded uniformly in $n$.
Hence  assertion (ii) follows from Proposition \ref{p:*}.
Moreover, it follows from the above argument that to prove assertion (i)  it suffices to estimate
\begin{equation}\label{int}
I_1:=(g\psi^{(n)}_{n-1},\epsilon(f\psi^{(n)}_{n})),\;
I_2:=(g\psi^{(n)}_{n-1},\epsilon(f\psi^{(n)}_{n-1})),\;
I_3:=(g\psi^{(n)}_{n},\epsilon(f\psi^{(n)}_{n})).
\end{equation}
with differentiable $f,g$.
  It follows from Proposition \ref{p:*} that
  \[I_1=I_{1,0}+\sum_{\alpha=1}^{2q}I_{1,\alpha}
+(g\psi^{(n)}_{n-1},\,\epsilon r_n)+O(n^{-1}),\]
where
\begin{eqnarray*}
I_{1,0}&=&n^{-1}k_n\int_{\sigma_{-\delta_n}}f(\lambda)g(\lambda)R_0(\lambda)R_1(\lambda)
\frac{\sin nF_n(\lambda)\sin nF_{n-1}(\lambda)}
{F'_n(\lambda)}d\lambda \\
I_{1,2\alpha}&=&
n^{-1/3}B^{(2\alpha)}_{11}B^{(2\alpha)}_{21}f(E_{2\alpha})g(E_{2\alpha})\int_{E_{2\alpha}-\delta_n}^{E_{2\alpha}+\delta_n}
\Psi\left(n^{2/3}\Phi_{2\alpha}(\lambda-E_{2\alpha})\right)
\\&&
\cdot\frac{Ai\left(n^{2/3}\Phi_{2\alpha}(\lambda-E_{2\alpha})\right)}{\Phi_{2\alpha}'(0)}d\lambda,
\end{eqnarray*}
and $I_{1,2\alpha-1}$ is the integral similar to $I_{1,2\alpha}$ for the region $|\lambda-E_{2\alpha-1}|\le\delta_n$.
It is easy to see that
\begin{eqnarray*}
I_{1,\alpha}&=&B^{(\alpha)}_{11}B^{(\alpha)}_{21}\frac{f(E_{\alpha})g(E_{\alpha})}
{2n(\Phi_{\alpha}'(0))^2}\,\left(1+o(1)\right),
\end{eqnarray*}
Moreover, (\ref{b_er}) and (ii) of Lemma \ref{l:2} yield
\[(g\psi^{(n)}_{n-1},\epsilon r_n)=-(\epsilon(g\psi^{(n)}_{n-1}), r_n)\le
\int|\epsilon(g\psi^{(n)}_{n-1})|\,|r_n|\,d\lambda=O(n^{-1-3\kappa/4}).\]
Hence  we are left to find the bound for $I_{1,0}$.
\begin{eqnarray*}
I_{1,0}&=&(2n)^{-1}\int_{\sigma_{-\delta_n}}f(\lambda)g(\lambda)R_0(\lambda)R_1(\lambda)
\frac{\cos\pi(m_0(\lambda)-m_{1}(\lambda))}
{F'_n(\lambda)}d\lambda\\&&
+(2n)^{-1}\int_{\sigma_{-\delta_n}}f(\lambda)g(\lambda)R_0(\lambda)R_1(\lambda)
\frac{\cos n(F_n(\lambda)+F_{n-1}(\lambda))}
{F'_n(\lambda)}d\lambda=I_{10}'+I_{10}''.
\end{eqnarray*}
By (\ref{rel_R}) and (\ref{F_n}), we obtain
\[I_{11}'=\frac{1+o(1)}{2n}\int_{-2}^{2}\frac{f(\lambda)g(\lambda)}
{P(\lambda)X^{1/2}(\lambda)}d\lambda.\]
Integrating by parts, one can obtain easily that $I_{11}''=O(n^{-2}\delta_n^{-3/2})=O(n^{-1-3\kappa/2})$.

The other two integrals from (\ref{int}) can be estimated similarly.

\medskip

 {\bf Acknowledgements.} The author  thanks  Prof. T. Kriecherbauer
 and  Prof. B. Eynard for the  fruitful discussion.

\end{document}